\documentclass[aps,twocolumn,preprintnumbers,amsmath,amssymb,nofootinbib,superscriptaddress,notitlepage]{revtex4-1}

\usepackage{slashed}
\usepackage{soul}
\usepackage{epsfig}
\usepackage{mathrsfs,bm}
\usepackage{txfonts}
\usepackage{amssymb}
\usepackage{indentfirst}
\usepackage{graphicx,booktabs}
\usepackage{multirow}
\usepackage{overpic}
\usepackage{color}
\usepackage{amssymb}

\usepackage[utf8]{inputenc}

\newcommand{\corr}[1]{{#1}}

\begin{document}
\title{Molecular $P_{\psi}$ pentaquarks from light-meson exchange saturation}

\author{Zi-Ying Yang}
\affiliation{School of Physics,  Beihang University, Beijing 100191, China}

\author{Fang-Zheng Peng}
\affiliation{Southern Center for Nuclear-Science Theory (SCNT), Institute of Modern Physics, Chinese Academy of Sciences, Huizhou 516000, China}%
\affiliation{Institute of Modern Physics, Chinese Academy of Sciences, Lanzhou 730000, China}%

\author{Mao-Jun Yan}
\affiliation{CAS Key Laboratory of Theoretical Physics, 
  Institute of Theoretical Physics,
  Chinese Academy of Sciences, Beijing 100190}
\affiliation{School of Physical Science and Technology, Southwest University,
  Chongqing 400715, China}

\author{Mario S\'anchez S\'anchez}
\affiliation{LP2IB (CNRS/IN2P3 – Universit\'e de Bordeaux), 33175 Gradignan cedex, France}
\affiliation{Departamento de Física, Universidad de Murcia, 30071 Murcia, Spain}

\author{Manuel Pavon Valderrama}\email{mpavon@buaa.edu.cn}
\affiliation{School of Physics,  Beihang University, Beijing 100191, China}

\date{\today}
\begin{abstract}
  Theoretical predictions of the spectrum of heavy meson-baryon bound states
  are a fundamental tool for disentangling the nature of the different
  pentaquark states that have been observed in experimental
  facilities.
  Here we explore this spectrum in a phenomenological model that describes
  the heavy meson-baryon interaction in terms of a contact-range
  interaction, where the coupling strength is saturated by
  the exchange of light scalar and vector mesons,
  i.e. $\sigma$, $\rho$ and $\omega$ exchanges.
  Saturation determines the couplings modulo an unknown proportionality
  constant that can be calibrated from a molecular candidate. 
  If we use the $P_{\psi}^N(4312)$ as input, we predict a series of molecular
  pentaquarks including the $P_{\psi}^N(4440)$ and $P_{\psi}^N(4457)$,
  the recent $P_{\psi s}^{\Lambda}(4338)$ and
  the $P_{\psi s}^{\Lambda}(4459)$.
\end{abstract}

\maketitle

\section{Introduction}

Recently the LHCb collaboration has observed~\cite{LHCb:2022jad} 
a new pentaquark state within the $J/\psi \Lambda$ invariant mass distribution.
Its mass and width (in units of ${\rm MeV}$) are
\begin{eqnarray}
  M &=& 4338.2 \pm 0.7  \, , \quad
  \Gamma = 7.0 \pm 1.2 \, , \label{eq:ms1} 
\end{eqnarray}
and its spin-parity is $J^P = \tfrac{1}{2}^-$.
This $P_{\psi s}^{\Lambda}(4338)$ pentaquark is simply the latest addition
to a growing family of hidden-charm pentaquark states, which includes
the $P_{\psi s}^{\Lambda}(4459)$~\cite{Aaij:2020gdg},
also discovered in $J/\psi \Lambda$ and
with mass and width (${\rm MeV}$)
\begin{eqnarray}
  M &=&
  4458.8 \pm 2.9 {}^{+4.7}_{-1.1} \, , \quad
  \Gamma = 17.3 \pm 6.5 {}^{+8.0}_{-5.7} \, , \label{eq:ms2}  
\end{eqnarray}
or the three $P_{\psi}^N$ states discovered in 2019~\cite{Aaij:2019vzc} ---
the $P_{\psi}^N(4312)$, $P_{\psi}^N(4440)$ and $P_{\psi}^N(4457)$ ---
whose masses and widths are, respectively (again in ${\rm MeV}$):
\begin{eqnarray}
  M &=&4311.9\pm 0.7^{+6.8}_{-0.6} \, ,  \quad
  \Gamma =9.8\pm2.7^{+3.7}_{-4.5} \, , \label{eq:m1} \\
  M &=&4440.3\pm 1.3^{+4.1}_{-4.7} \, , \quad
  \Gamma =20.6\pm4.9^{+8.7}_{-10.1} \, , \label{eq:m2} \\
  M &=&4457.3\pm 0.6{}^{+4.1}_{-1.7} \, , \quad
  \Gamma =6.4\pm2.0^{+5.7}_{-1.9} \, . \label{eq:m3}
\end{eqnarray}
The masses of the aforementioned pentaquarks are all close to charmed
antimeson - charmed baryon thresholds and have thus been theorized to
be molecular, where specific discussions can be found
in~\cite{Karliner:2022erb,Wang:2022tmp,Yan:2022wuz,Meng:2022wgl,Burns:2022uha}
for the $P_{\psi s}^{\Lambda}(4338)$,
in~\cite{Chen:2020uif,Peng:2020hql,Liu:2020hcv,Chen:2020kco}
for the $P_{\psi s}^{\Lambda}(4459)$
and in~\cite{Chen:2019asm,Chen:2019bip,Liu:2019tjn,Xiao:2019aya,Valderrama:2019chc,Liu:2019zvb,Guo:2019fdo,Pan:2019skd}
for the $P_{\psi}^{N}(4312/4440/4457)$.
Yet, there are also non-molecular explanations~\cite{Eides:2019tgv,Stancu:2020paw,Ferretti:2020ewe,Ferretti:2021zis,Garcilazo:2022kra,Ortega:2022uyu}
and at least one pentaquark
--- the $P_{\psi}^{N}(4337)$~\cite{Aaij:2021august} ---
that is more difficult to accommodate
within the molecular picture~\cite{Yan:2021nio,Nakamura:2021dix}.

Here we will consider the previous five pentaquarks
from a molecular perspective~\footnote{By {\it molecular} we specifically
  refer to composite in the sense that most of the wave function of
  the state corresponds to a meson-baryon component
  (instead of a multiquark component).
}.
We are interested in the question of whether they can be explained
with the same set of parameters in the phenomenological model
we proposed in~\cite{Peng:2021hkr}.
This model describes molecular states in terms of a simple contact-range
interaction in which the exchange of light-mesons --- the scalar $\sigma$ and
the vector $\rho$ and $\omega$ mesons --- saturates the couplings.
The rationale behind the choice of a contact-range theory is that
the size of most hadronic bound states is not small enough
to probe the details of the light-meson exchange potential,
hence the simplification of the interaction from finite- to contact-range.
A few of these pentaquarks have indeed been explained
(or even predicted~\cite{Wu:2010vk,Wu:2010rv,Wang:2011rga}) by the exchange of
light-mesons~\cite{Xiao:2013yca,Xiao:2019gjd,Dong:2021juy}.
Yet, these models usually (though not always~\cite{Dong:2021juy})
concentrate on a particular sector (non-strange/strange).
By treating the non-strange and strange sectors on the same footing,
we will be able to answer the question of whether the $P_{\psi}^N$ and
$P_{\psi s}^{\Lambda}$ pentaquarks are indeed connected.

\section{Saturation of the contact-range couplings}
In the molecular picture the hidden-charmed pentaquarks are
charmed antimeson - charmed baryon two-body systems.
We describe the antimeson-baryon interaction in terms of a non-relativistic
S-wave contact-range potential of the form
\begin{eqnarray}
  \langle \vec{p}\,' | V_C | \vec{p} \rangle =
  C_0 + C_1\,\hat{\vec{S}}_{L1} \cdot \hat{\vec{S}}_{L2} \, , 
\end{eqnarray}
where there is a central and spin-spin piece, with $C_0$ and $C_1$ their
respective couplings, and with $\hat{\vec{S}}_{Li} = {\vec{S}}_{Li} / | S_{Li} |$
the normalized light-quark spin operator of hadron $i = 1,2$ (with $i = 1$
the antimeson and $i=2$ the baryon).
Being a contact-range interaction, this potential has to be regularized,
which we will do by introducing a regularization scale $\Lambda$,
i.e. a cutoff. Details on the specific regularization procedure
will be given later on.

This description is expected to be a good approximation to the baryon-antimeson
potential provided the two following conditions are met:
(i) the binding momentum of the molecular pentaquarks is smaller than
the mass of the light-mesons generating the binding, which we suspect
to be the scalar and vector mesons, and (ii) that long-range effects,
such as one pion exchange (OPE) are perturbative.
The first of these conditions is met for the molecular interpretations of
the $P_{\psi}^N(4312/4440/4457)$ and $P_{\psi s}^{\Lambda}(4338/4459)$
pentaquarks, i.e. $\bar{D}^{(*)} \Sigma_c$ and $\bar{D}^{(*)} \Xi_c$
bound states. Indeed, the binding momentum of these systems
(defined as $\gamma = \sqrt{2 \mu B}$ with $\mu$ the reduced mass of the
two-body system and $B$ the binding energy) is of the order of
$(100-200)\,{\rm MeV}$, while the $\rho$ meson mass
is about $770\,{\rm MeV}$.
The second of these conditions is trivially met for the $\bar{D} \Sigma_c$
and $\bar{D}^{(*)} \Xi_c$ systems, for which OPE is forbidden,
while for the $\bar{D}^* \Sigma_c$ case
there are EFT arguments~\cite{Valderrama:2019chc}
and explicit calculations in the phenomenological model
we are using here~\cite{Peng:2021hkr}
that show that OPE is indeed a perturbative correction and
can be safely neglected in a first approximation.

The couplings $C_0$ and $C_1$ are assumed to be saturated by scalar and
vector meson exchange. We first remind that the potentials generated
by the scalar and vector light-mesons are
\begin{eqnarray}
  V_S &=& - \frac{g_{S1} g_{S2}}{m_S^2 + {\vec{q}\,}^2} \, ,
  \label{eq:V_S} \\
  V_V &=& (\zeta + T_{12}) \, \Big[ \frac{g_{V1} g_{V2}}{m_V^2 + {\vec{q}\,}^2}
  + \frac{f_{V1} f_{V2}}{6 M^2}\,\frac{m_V^2}{m_V^2 + {\vec{q}\,}^2}\,
  \hat{\vec{S}}_{L1} \cdot \hat{\vec{S}}_{L2} \Big] \nonumber \\
  && + \dots \, , \label{eq:V_V}
\end{eqnarray}
where $m_S$ and $m_V$ are the scalar and vector meson masses, $g_{Si}$,
$g_{Vi}$, and $f_{Vi}$ are coupling constants (with $i=1,2$ indicating
whether we are referring to the charmed antimeson or baryon),
$M$ a scaling mass, which we take to be $M = m_N$ with
$m_N = 938.9\,{\rm MeV}$ the nucleon mass, and the dots
represent either Dirac-deltas or higher partial
wave contributions to the potential.
The vector meson potential has been written for the particular case
in which only the $\rho$ and $\omega$ mesons are exchanged, i.e.
for the $\bar{D}^{(*)} \Sigma_c^{(*)}$ and
$\bar{D}^{(*)} \Xi_c^{('/*)}$ systems,
where $\zeta = \pm 1$ is the sign of $\omega$ exchange
($\zeta = +1$ for hidden-charm pentaquarks and
$\zeta = -1$ for doubly charmed ones) and
$T_{12}$ is an isospin factor given by
$T_{12} = \vec{\tau}_1 \cdot \vec{T}_2$ for $\bar{D}^{(*)} \Sigma_c^{(*)}$
and $T_{12} = \vec{\tau}_1 \cdot \vec{\tau}_2$ for $\bar{D}^{(*)} \Xi_c^{('/*)}$,
{where we remind that $\vec{\tau}_i$ ($\vec{T}_i$) are the Pauli
  (spin-1) matrices when acting on the isospin degrees of freedom of
  an isospin-$1/2$ (isospin-$1$) hadron}.
For the masses of the light mesons we take $m_S = 475\,{\rm MeV}$ (i.e.
the average of its $(400-550)\,{\rm MeV}$ mass range
in the Review of Particle Physics (RPP)~\cite{Workman:2022ynf}) and
$m_V = (m_\rho + m_{\omega})/2 = 775\,{\rm MeV}$
(i.e. the average of the $\rho$ and $\omega$ masses).
For the scalar meson couplings, we use
the linear sigma model~\cite{GellMann:1960np} and
the quark model~\cite{Riska:2000gd}
plus the assumption that the coupling of the scalar meson to the $u$, $d$ and
$s$ quarks is the same, yielding $g_{S} = 3.4$ and $6.8$
for the charmed mesons and baryons, respectively.
For the vector meson couplings we use Sakurai's universality and vector meson
dominance~\cite{Sakurai:1960ju,Kawarabayashi:1966kd,Riazuddin:1966sw} to
obtain $g_V = 2.9$ and $5.8$ for the charmed mesons and $\Sigma_c^{(*)}$
baryons and $g_V = 2.9$ for the $\Xi_c^{('/*)}$ baryons;
for the magnetic-like couplings we define
$f_{V} = \kappa_V g_V$ with $\kappa_V = \frac{3}{2}\,({\mu_u}/{\mu_N})$
for both charmed mesons and sextet charmed baryons
($\Sigma_c$, $\Sigma_c^*$, $\Xi_c'$, $\Xi_c^*$),
with $\mu_N$ the nuclear magneton and $\mu_u \simeq 1.9\,\mu_N$ the magnetic
moment of a constituent u-quark. In contrast, for antitriplet charmed baryons
($\Xi_c$) we have $\kappa_V = 0$ instead.
The extension to the other molecular configurations can be done by modifying
the vector meson masses to take into account that the $K^*$ and $\phi$ vector
mesons are heavier than the $\rho$ and $\omega$ and
by changing the $(\zeta + T_{12})$ factor
by the corresponding SU(3)-flavor factor
(we will explain these changes
in more detail later on).

Provided we use a regularization scale of the same order of magnitude
as the mass of the light mesons, the saturation of the couplings
by a scalar meson will be given by
\begin{eqnarray}
  C_0^S(\Lambda \sim m_S) &\propto& - \frac{g_{S1} g_{S2}}{m_S^2} \, , \\
  C_1^S(\Lambda \sim m_S) &\propto& 0 \, , 
\end{eqnarray}
while the saturation by a vector meson will be
\begin{eqnarray}
  C_0^V(\Lambda \sim m_V) &\propto&
  \frac{g_{V1} g_{V2}}{m_V^2}\,(\zeta + T_{12}) \, , \\
  C_1^V(\Lambda \sim m_V) &\propto&
  \frac{f_{V1} f_{V2}}{6 M^2}\,(\zeta + T_{12}) \, ,
\end{eqnarray}
where the proportionality constant is, in principle unknown, and
must be determined from external information.
In the particular case of vector meson exchange we have followed
the novel saturation method proposed in~\cite{Peng:2020xrf}.

Owing to $m_S \neq m_V$, it is apparent that the ideal regularization scale
for the saturation of the $C_0$ coupling by the scalar and vector mesons
is not identical.
This is dealt with by considering the renormalization group evolution (RGE)
of the contact-range potential $V_C$,
which is given by~\cite{PavonValderrama:2014zeq}
\begin{eqnarray}
  \frac{d}{d \Lambda}\, \langle \Psi | V_C(\Lambda) | \Psi \rangle = 0 \, ,
\end{eqnarray}
where $\Psi$ represents the wave function of the two-body bound state under
consideration and $\Lambda$ the regularization scale.
If the r-space wave function displays a power-law behavior of the type
$\Psi(r) \sim r^{\alpha/2}$ at distances $r \sim 1/\Lambda$,
the previous RGE becomes
\begin{eqnarray}
  \frac{d}{d \Lambda}\,\left[ \frac{C(\Lambda)}{\Lambda^{\alpha}} \right] = 0
  \, , \label{eq:C-RGE}
\end{eqnarray}
where
$C(\Lambda) = C_0(\Lambda) + C_1(\Lambda)\,\hat{\vec{S}}_{L1} \cdot \hat{\vec{S}}_{L2}$.
This implies the following relation of the couplings when evaluated at
different regularization scales:
\begin{eqnarray}
  \frac{C(\Lambda_1)}{\Lambda_1^{\alpha}} =
  \frac{C(\Lambda_2)}{\Lambda_2^{\alpha}} \, . \label{eq:C-scaling}
\end{eqnarray}
That is, provided $\alpha$ is known, it will be trivial to combine
the contributions from scalar and vector meson exchange to
the saturation of the couplings.
In particular, if we fix the regularization scale to the vector meson mass,
we will have
\begin{eqnarray}
  C(\Lambda = m_V)
  &=& C_V(m_V) + {\left(\frac{m_V}{m_S}\right)}^{\alpha} \, C_S(m_S) \nonumber \\
  &\propto& 
  \frac{g_{V1} g_{V2}}{m_V^2}
  \left[ \zeta + {T}_{12} \right]
  \left( 1 + \kappa_{V1} \kappa_{V2} \, \frac{m_V^2}{6 M^2}\,\hat{C}_{L12} \right)
  \nonumber \\
  && \quad \,\, - {(\frac{m_V}{m_S})}^{\alpha}\frac{g_{S1} g_{S2}}{m_S^2} \, ,
    \label{eq:coupling-sat}
\end{eqnarray}
where $\hat{C}_{L12} = \hat{\vec{S}}_{L1} \cdot \hat{\vec{S}}_{L2}$.

The only problem is determining the scaling power $\alpha$.
{
  From the regularity of the wave function at the origin, we expect
  $\Psi(r) \sim 1$ for $r \to 0$, yielding $\alpha = 0$.
  But this is only the case for point-like particles:
  the finite size of hadrons will suppress the two-body
  wave functions at distances of the order of the hadron size
  (i.e. the molecular probability decreases, while the probability that
  the state is in a multiquark configuration increases).
  This observation implies that $\alpha > 0$.
  For instance, if we include a Lorentzian form factor in the momentum
  space representation of the wave function, i.e. a factor of the type
  $1/(1 + (q/M)^2) \to M^2/q^2$ for $q \gg M$ with $M$ the momentum
  scale at which the structure of hadrons becomes evident,
  this translates into an $(M r)^2$ suppression
  at distances $(M r) \ll 1$,
  suggesting $\alpha = 2$.
  Yet, scalar and vector meson exchange still involve slightly lower momenta
  than the ones for which the internal structure of the hadrons
  becomes evident ($M \sim 1\,{\rm GeV}$).
  Thus we will choose $\alpha = 1$, which takes into account that
  meson-exchanges happen in a region where the internal
  structure of hadrons is not fully resolved, yet
  considering them as point-like is not a good
  approximation either~\footnote{A more practical reason is that
    with $\alpha = 1$ the results of the RG-improved saturation
    method presented are basically the same as the ones obtained with the
    one-boson-exchange model for the molecular $\Sigma_c^{(*)} \bar{D}^{(*)}$
    pentaquarks when both methods are calibrated to reproduce
    the mass of the $P_{\psi}^N(4312)$ state~\cite{Liu:2019zvb}.}.
}

It is worth noticing that though we have expressed $C(\Lambda)$ for
$\Lambda = m_V$, it is perfectly acceptable to vary $\Lambda$
around this value: from Eqs.~(\ref{eq:C-RGE},\ref{eq:C-scaling})
we see that changing the regularization scale will simply change
the proportionality constant in the second line of Eq.~(\ref{eq:coupling-sat}).
Yet, we remind that we are dealing with a phenomenological model (and not
with an effective field theory): ultimately the value of $\Lambda$
is a parameter of the model and the criterion by which
it is chosen is the reproduction of known features of
the pentaquark spectrum {(an observation which
  is also applicable to the parameter $\alpha$
  we discussed in the previous paragraph)}.

\section{Calibration and predictions}

\subsection{Basic formalism}

For predictions to be made in this model, we first have to determine
the unknown proportionality constant in Eq.~(\ref{eq:coupling-sat}).
This is done by choosing a {\it reference state} --- a state that is assumed
to be a particular type of two-body bound state --- and then calculating
the strength of the coupling constant for this state.
Our choice of a reference state will be the $P_{\psi}^{N}(4312)$ pentaquark
as a $\Sigma_c \bar{D}$ bound state.

The way this is done is straightforward, though it involves a few steps.
First, we begin by regularizing the contact-range potential with a specific
separable regulator function $f(x)$
\begin{eqnarray}
  V_C = C^{\rm sat}_{\rm mol}(\Lambda)\,
  f(\frac{p'}{\Lambda}) f(\frac{p}{\Lambda}) \, ,
\end{eqnarray}
for which we will choose a Gaussian, $f(x) = e^{-x^2}$ and where
$C^{\rm sat}_{\rm mol}$ is the saturated coupling
for a particular molecular state ``${\rm mol}$''.
The regularization scale will be taken to be $\Lambda = 1.0\,{\rm GeV}$,
i.e. the value we already used in~\cite{Peng:2021hkr}.
The regularized potential is then inserted in the Lippmann-Schwinger equation
for the bound state pole, which for a separable interaction takes the
simplified form
\begin{eqnarray}
  1 + C^{\rm sat}_{\rm mol}(\Lambda)\,
  \int \frac{d^3 \vec{q}}{(2 \pi)^3}\,
  \frac{f^2(\frac{q}{\Lambda})}
       {M_{\rm th} + \frac{q^2}{2 \mu_{\rm mol}} - M_{\rm mol}}
   = 0\, , \nonumber \\ \label{eq:single-channel-general}
\end{eqnarray}
which implicitly assumes that we are dealing with a single-channel problem.
Here $M_{\rm mol}$ refers to the mass of the molecule (or pole of the T-matrix),
$M_{\rm th} = m_1 + m_2$ to the mass of the two-body threshold and
$\mu = m_1 m_2 / (m_1 + m_2)$ to the reduced mass of the two-body system,
with $m_1$ and $m_2$ the masses of particles $1$ and $2$.
For the masses of the charmed antimesons and charmed baryons we will take
the isospin average of the values listed in the RPP~\cite{Workman:2022ynf}.

By particularizing the previous equation for a reference state
(${\rm mol} = {\rm ref}$), we obtain $C^{\rm sat}_{\rm ref}$.
For $P_{\psi}^N(4312)$ as the reference state, we find
$C^{\rm sat}_{\rm ref} = -0.80^{+0.14}_{-0.01}\,{\rm fm}^2$
for $\Lambda = 1.0\,{\rm GeV}$, where the errors come from
the uncertainties in the mass of the $P_{\psi}^N(4312)$.
From this, now we can easily solve the bound state equation for any other
(single-channel) molecule simply by making the substitution:
\begin{eqnarray}
  1 + 2\mu_{\rm ref}\,C^{\rm sat}_{\rm ref}(\Lambda)\,R_{\rm mol}\,
  \int \frac{d^3 \vec{q}}{(2 \pi)^3}\,
  \frac{f^2(\frac{q}{\Lambda})}
       {q^2 + \gamma_{\rm mol}^2}
   = 0\, , \nonumber \\ \label{eq:single-channel}
\end{eqnarray}
where $\gamma_{\rm mol} = \sqrt{2 \mu_{\rm mol}\,(M_{\rm th} - M_{\rm mol})}$ is
the wave number of the bound state, and $R_{\rm mol}$ is the ratio between
the relative strength of the coupling for the molecule ``mol'' and
the reference state:
\begin{eqnarray}
  R_{\rm mol} =
  \frac{\mu_{\rm mol}\,C^{\rm sat}_{\rm mol}}{\mu_{\rm ref}\,C^{\rm sat}_{\rm ref}} \, ,
  \label{eq:Rmol}
\end{eqnarray}
which is independent of the unknown proportionality constant
in Eq.~(\ref{eq:coupling-sat}).
It is interesting to notice that for our choice of regulator ($f(x) = e^{-x^2}$)
the loop integral is analytic and given by
\begin{eqnarray}
         && \int \frac{d^3 \vec{q}}{(2 \pi)^3}\,
  \frac{f^2(\frac{q}{\Lambda})}
       {q^2 + \gamma_{\rm mol}^2} = \nonumber \\
       &&  \qquad \frac{1}{8 \pi^2}\,\left[
         \sqrt{2 \pi}\,\Lambda -
    2\, e^{2 \gamma_{\rm mol}^2 / \Lambda^2}\,\pi \gamma_{\rm mol} \,
    {\rm erfc}\left( \frac{\sqrt{2} \gamma_{\rm mol}}{\Lambda} \right)
         \right]  \, , \nonumber \\ \label{eq:loop-gaussian}
\end{eqnarray}
with ${\rm erfc}\,(x)$ the complementary error function.
Solutions for which $\gamma_{\rm mol} > 0$ and $\gamma_{\rm mol} < 0$
correspond to bound and virtual states, respectively.

\begin{figure}
  \begin{center}
    \epsfig{figure=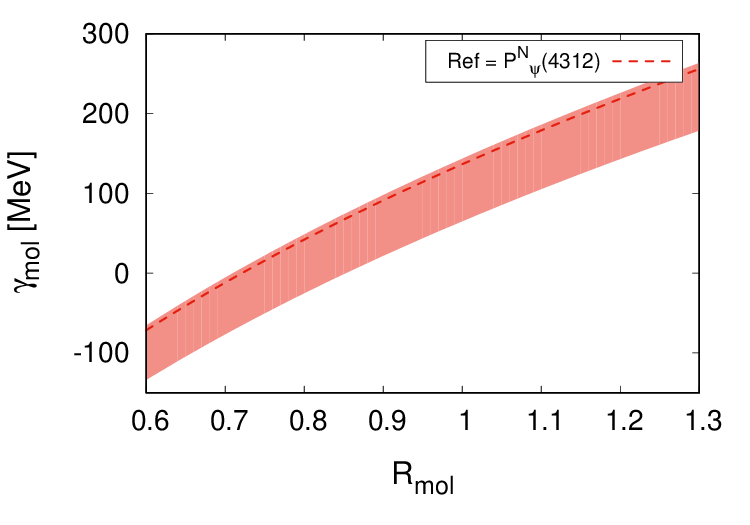,
      width=8.5cm}
    \end{center}
  \caption{
    Dependence of the wave number of a molecule ($\gamma_{\rm mol}$,
    in ${\rm MeV}$) on the molecular ratio $R_{\rm mol}$
    (defined in Eq.~(\ref{eq:Rmol}))
    when the reference state is the $P_{\psi}^N(4312)$.
    This dependence is calculated by solving
    Eq.~(\ref{eq:single-channel}) with $\Lambda = 1.0\,{\rm GeV}$,
    $C^{\rm sat}_{\rm ref} = -0.80^{+0.14}_{-0.01}\,{\rm fm}^2$ (given by reproducing
    the mass of the $P_{\psi}^N(4312)$) and $R_{\rm mol}$ as inputs.
    The band represents the uncertainty coming
    from the mass of the reference state.
    Values of $\gamma_{\rm mol} > 0$ ($\gamma_{\rm mol} < 0$) indicate
    the prediction of a bound (virtual) state.
  }
\label{fig:gamma-mol}
\end{figure}

Predictions of $\gamma_{\rm mol}$ only depend on the ratio $R_{\rm mol}$:
$\gamma_{\rm mol} = \gamma_{\rm mol}(R_{\rm mol})$.
We show this numerical dependence in Fig.~\ref{fig:gamma-mol}
for our cutoff and reference state choices.
Bound (virtual) state solutions require:
\begin{eqnarray}
  R_{\rm mol} > 0.72^{+0.13}_{-0.01} \quad (R_{\rm mol} \leq 0.72^{+0.13}_{-0.01}) \, ,
\end{eqnarray}
where we remind that $R_{\rm mol}$ is defined relative to the reference
state $P_{\psi}^N(4312)$ (i.e. other choices of input state will
result in different values).

\subsection{Error estimations}

We consider {four} error sources for the calculation of the masses of
meson-baryon states:
\begin{itemize}
\item[(i)] the uncertainty of the mass of the reference state, 
  the $P_{\psi}^N(4312)$, which yields
  $C^{\rm sat}_{\rm ref} = -0.80^{+0.14}_{-0.01}\,{\rm fm}^2$,
\item[(ii)] the uncertainty of the mass of the scalar meson, for which
  we take $m_S = (475 \pm 75)\,{\rm MeV}$, i.e., equivalent to
  the $m_S = (400-550)\,{\rm MeV}$ range listed
  in the RPP~\cite{Workman:2022ynf},
\item[(iii)] {the choice of a regularization scale, for which we vary
  $\Lambda$ in the $(0.75-1.5)\,{\rm GeV}$ range, where the central
  value we use --- i.e. $\Lambda = 1.0\,{\rm GeV}$ --- is the harmonic mean of
  the lower and upper bounds of said range~\footnote{This assumes that finite cutoff effects
    scale as $1/\Lambda$, hence the use of the harmonic mean. Yet,
    strictly speaking this is only true for a renormalizable
    contact-range theory, which is not the case, as we are
    dealing with a phenomenological model that requires
    $\Lambda$ to be of the order of the vector
    meson mass.}, and.
\item[(iv)] {the choice of the parameter $\alpha$, for which we take
  $\alpha = 1 \pm 1$. This parameter accounts for the effects of
  the finite size of the hadrons, with $\alpha = 0$ corresponding to
  point-like hadrons, while $\alpha > 0$ mimics their finite size.}
}
\end{itemize}
{We notice that the first two sources of uncertainty are external
  to the model, while the last two correspond to the possible choices
  of the internal parameters of the model.
}

For the final error we choose the largest of these sources:
\begin{eqnarray}
  \Delta M_{\rm mol} &=& {\rm minmax}\,\Big\{
  M_{\rm mol}(C^{\rm sat}_{\rm ref} \pm \Delta C^{\rm sat}_{\rm ref}) - M_{\rm mol} \, ,
  \nonumber \\
  && \qquad  \qquad M_{\rm mol}(m_S \pm \Delta m_S) - M_{\rm mol} \, ,
  \nonumber \\
  && \qquad \qquad M_{\rm mol}(\Lambda \pm \Delta \Lambda) - M_{\rm mol} \, ,
  \nonumber \\
  && \qquad \qquad M_{\rm mol}(\alpha \pm \Delta \alpha) - M_{\rm mol} 
  \Big\} \, ,
\end{eqnarray}
with $M_{\rm mol}$ the central value of the prediction and where
``${\rm minmax}$'' refers to the maximum (for the positive error)
and minimum (for the negative error) number within the list.

\subsection{Coupled channel dynamics}

For the molecular pentaquarks in which coupled channel dynamics are
required, which happens when there are two nearby meson-baryon thresholds
with the same quantum numbers, we will use the previous equation
in matrix form
\begin{eqnarray}
    \phi_A + \sum_{B} \phi_B\,C^{\rm sat}_{\rm mol (AB)}(\Lambda)\,\int \frac{d^3 \vec{q}}{(2 \pi)^3}\,
  \frac{f^2(\frac{q}{\Lambda})}{M_{\rm th(B)} + \frac{q^2}{2 \mu_B} - M_{\rm mol}}
   = 0\, , \nonumber \\
\end{eqnarray}
where $A$, $B$ are labels denoting the meson-baryon channels, $M_{\rm th(B)}$
is the mass of the meson-baryon channel $B$ and $\phi_A$, $\phi_B$ are
the vertex functions for channels $A$, $B$: $\phi_A$ is related to
the wave function of channel $A$ by means of the relation
$({M_{\rm mol} - \frac{q^2}{2 \mu} - M_{\rm th(A)}}) \Psi_A(\vec{q}) = \phi_A \,
f(\frac{q}{\Lambda})$.
The vertex function $\phi_A$ is thus related to the coupling of a molecular
pentaquark to the meson-baryon channel $A$.
If we write the equations in terms of the relative strength
$R_{\rm mol}^{AB} = \mu_{\rm mol (B)} C_{\rm mol}^{\rm sat (AB)} / (\mu_{\rm ref} C_{\rm ref}^{\rm sat})$, we have
\begin{eqnarray}
\phi_A + 2\mu_{\rm ref} \sum_{B} \phi_B\,C^{\rm sat}_{\rm ref}(\Lambda)\,R_{\rm mol}^{AB}\,\int \frac{d^3 \vec{q}}{(2 \pi)^3}\,
\frac{f^2(\frac{q}{\Lambda})}{q^2 + \gamma_B^2}
= 0\, , \nonumber \\
\end{eqnarray}
where $\gamma_B = \sqrt{2 \mu_B (M_{\rm th(B)} - M_{\rm mol})}$.
For the Gaussian regulator the loop integral is again
given by Eq.~(\ref{eq:loop-gaussian}).
The coupled channel equations admit solutions with different signs of
$\gamma_A$, which we will characterize as being in Riemann sheets
I ($\gamma_A > 0$) and II ($\gamma_A < 0$).

There are two types of coupled channel dynamics to be considered, namely
\begin{itemize}
\item[(i)] stemming from heavy-quark spin symmetry (HQSS) and,
\item[(ii)] stemming from SU(3)-flavor symmetry.
\end{itemize}
The first type involves the transition of a ground state heavy hadron into
an excited state one, e.g., $\bar{D} \Sigma_c$-$\bar{D}^* \Sigma_c$ or
$\bar{D}^* \Sigma_c$-$\bar{D}^* \Sigma_c^*$.
Within our model this type of dynamics requires the spin-spin or $M1$ component
of vector meson exchange, which is weaker than the central or $E0$ component.
Besides, the mass splittings involved are usually large.
This leads in general to relatively modest effects from HQSS coupled channel
dynamics, a point that was explicitly checked with concrete calculations
for the pentaquarks in Ref.~\cite{Peng:2021hkr}.
We will thus ignore this type of coupled channel dynamics.

The second type involves a transition between two different meson-baryon
channels for which there is a common irreducible component of their
SU(3)-flavor decomposition (e.g., both channels contain
an octet component).
To understand this point better, we begin by considering the two types of
molecular pentaquarks appearing in this work, the ones containing
antitriplet charmed mesons ($T_c = \Lambda_c, \Xi_c$) and
the ones containing sextet charmed mesons
($S_c = \Sigma_c, \Xi_c', \Omega_c$ and $\Sigma_c^*, \Xi_c^*, \Omega_c^*$).
We refer to them by the notation $\bar{H}_c T_c$ and $\bar{H}_c S_c$, where
$\bar{H}_c = \bar{D}, \bar{D}_s$ and $\bar{D}^*, \bar{D}_s^*$.
From the point of view of SU(3)-flavor symmetry, the $\bar{H}_c T_c$ and
$\bar{H}_c S_c$ systems admit, respectively, the decompositions
$3 \otimes \bar{3} = 1 \oplus 8$ and $3 \otimes 6 = 8 \oplus 10$.
This will give rise to the following subtypes of coupled channel dynamics:
\begin{itemize}
\item[(a)] Most of the $\bar{H}_c S_c$ pentaquarks are pure octets
  or decuplets and do not mix. The exceptions are the isovector
  $\bar{D}_s \Sigma_c$ and $\bar{D} \Xi_c^{'}$ and isodoublet
  $\bar{D}_s \Xi_c^{'}$ and $\bar{D} \Omega_c$ channels, plus the corresponding
  channels involving the excited charmed antimesons and sextet baryons,
  for which we have~\cite{Kaeding:1995vq}:
  \begin{eqnarray}
    | \bar{D}_s \Sigma_c \rangle &=&
    + \sqrt{\frac{2}{3}}\,|8 \rangle + \frac{1}{\sqrt{3}}\,|10\rangle \, ,
    \label{eq:HS-decomp-1} \\ 
    | \bar{D} \Xi_c'(1) \rangle &=&
    - \frac{1}{\sqrt{3}}\,|8 \rangle + \sqrt{\frac{2}{3}}\,|10\rangle \, ,  \\
    \nonumber \\
    | \bar{D}_s \Xi_c' \rangle &=&
    + \frac{1}{\sqrt{3}}\,|8 \rangle + \sqrt{\frac{2}{3}}\,|10\rangle \, ,
    \\
    | \bar{D} \Omega_c \rangle &=&
    - \sqrt{\frac{2}{3}}\,|8 \rangle + \frac{1}{\sqrt{3}}\,|10\rangle \, ,
    \label{eq:HS-decomp-4} 
  \end{eqnarray}
  where the number in parentheses denotes the isospin
  (if there is any ambiguity) and $| 8 \rangle$ and $| 10 \rangle$
  denote the SU(3)-flavor irreducible representation.
  As a consequence, these channels will be coupled. The mass difference
  between the thresholds is of the order of $20\,{\rm MeV}$, which
  might justify the explicit inclusion of this effect,

\item[(b)] Most of the $\bar{H}_c T_c$ pentaquarks are pure octets, except
  for the $\bar{D}_s \Lambda_c$ and isoscalar $\bar{D} \Xi_c$ channels (and
  the channels involving the excited charmed antimesons), which are a
  mixture of singlet and octet~\cite{Kaeding:1995vq}:
  \begin{eqnarray}
    | \bar{D}_s \Lambda_c \rangle &=&
    \frac{1}{\sqrt{3}}\,| \tilde{1} \rangle + \sqrt{\frac{2}{3}}\,| \tilde{8} \rangle \, , 
    \\
    | \bar{D} \Xi_c(0) \rangle &=&
    \sqrt{\frac{2}{3}}\,| \tilde{1} \rangle - \frac{1}{\sqrt{3}}\,| \tilde{8} \rangle \, , 
  \end{eqnarray}
  where the notation is analogous to
  Eqs.~(\ref{eq:HS-decomp-1}-\ref{eq:HS-decomp-4}), though we have marked
  the SU(3)-flavor representations with a tilde to indicate that they are
  not the same as in the $\bar{H}_c S_c$ case.
  Again, we will have coupled channel effects. However, the mass difference
  between these thresholds is about $80\,{\rm MeV}$, suggesting this is less
  important than the $\bar{D} \Xi_c^{'}$-$\bar{D}_s \Sigma_c$ or
  $\bar{D}_s \Xi_c^{'}$-$\bar{D} \Omega_c$ coupled channel dynamics.

\item[(c)] Finally, the octet pieces of the $\bar{H}_c T_c$ and $\bar{H}_c S_c$
  pentaquarks can also generate mixing. In our model this effect requires
  the spin-spin / $M1$ component of vector meson exchange, and it is thus
  expected to be somewhat weaker when compared to the previously mentioned
  coupled channel dynamics. Yet, in a few instances the mass difference
  between the thresholds involved is not particularly large (e.g.
  the $\bar{D}^* \Xi_c$ channel with the $\bar{D} \Xi_c'$ or
  $\bar{D} \Xi_c^*$ channels), making this effect more
  important than naively expected. 
\end{itemize}
Based on the previous arguments, we expect a hierarchy of coupled channel
effects, with (a) being more important than (b) and (c).
%
For the particular case of (b) and (c) their relative size will however
depend on the specifics of the thresholds involved
(and their mass gaps).
Yet, concrete calculations in Ref.~\cite{Yan:2021nio} indicate
that the size of these effects is small except for a few
selected molecular configurations
(e.g. $\bar{D}^* \Xi_c (I=1)$-$\bar{D}_s \Sigma_c $).
%
%
In the following lines we will elaborate further by calculating the spectrum
first in a simplified approach where only type (a) of coupled channel dynamics
is included and then in a more complete approach also including
the type (b) and (c) dynamics when applicable.

\subsection{Predictions with minimal coupled channel dynamics}

\begin{table}[!ttt]
\begin{tabular}{|clllcl|}
\hline\hline
System  & $I$($J^{P}$) & $R_{\rm mol}$ & $M_{\rm mol}$ &
Candidate & $M_{\rm candidate}$\\
\hline
$\Lambda_c \bar{D}$ & $\tfrac{1}{2}$ ($\tfrac{1}{2}^{-}$) & $0.69$ & \corr{$(4153.6^{+0.1(B)}_{-4.9})^V$} & - & - \\
$\Lambda_c \bar{D}^*$ & $\tfrac{1}{2}$ ($\tfrac{1}{2}^{-}$,$\tfrac{3}{2}^{-}$) & $0.72$ & \corr{$(4295.0^{+0.0(B)}_{-3.6})^V$} & - & - \\
\hline
$\Lambda_c \bar{D}_s$ & $0$ ($\tfrac{1}{2}^{-}$) & $0.86$ & 
$4252.5^{+2.3}_{-2.0}$ & - & - \\
$\Lambda_c \bar{D}_s^*$ & $0$ ($\tfrac{1}{2}^{-}$,$\tfrac{3}{2}^{-}$) & $0.89$ &
$4395.2^{+3.2}_{-2.3}$ & - & - \\
\hline
$\Xi_c \bar{D}$ & $0$ ($\tfrac{1}{2}^{-}$) & $1.00$ & 
$4327.4^{+6.9}_{-0.9}$ & $P_{\psi s}^{\Lambda}(4338)$ & $4338.2 \pm 0.7$ \\
$\Xi_c \bar{D}^*$ & $0$ ($\tfrac{1}{2}^{-}$,$\tfrac{3}{2}^{-}$) & $1.04$ & 
\corr{$4466.7^{+7.8}_{-1.3}$} & $P_{\psi s}^{\Lambda}(4459)$ & $4458.9^{+5.5}_{-3.1}$ \\
$\Xi_c \bar{D}$ & $1$ ($\tfrac{1}{2}^{-}$) & $0.72$ & 
\corr{$(4336.3^{+0.0(B)}_{-3.8})^V$} & - & - \\
$\Xi_c \bar{D}^*$ & $1$ ($\tfrac{1}{2}^{-}$,$\tfrac{3}{2}^{-}$) & $0.74$ & 
\corr{$4477.6^{+0.0(V)}_{-2.4}$} & - & - \\
$\Xi_c \bar{D}_s$ & $\tfrac{1}{2}$ ($\tfrac{1}{2}^{-}$) & $0.82$ & 
\corr{$4436.3^{+1.2}_{-2.7}$} & - & - \\
$\Xi_c \bar{D}_s^*$ & $\tfrac{1}{2}$ ($\tfrac{1}{2}^{-}$,$\tfrac{3}{2}^{-}$) & $0.85$ & 
\corr{$4579.2^{+2.1(V)}_{-3.3}$} & - & - \\
\hline\hline
\end{tabular}
\caption{Molecular triplet-antitriplet pentaquark states predicted in this work.
  ``System'' indicates which charmed antimeson-baryon system
  we are dealing with,
  $I(J^P)$ the isospin, spin and parity of the molecule, $R_{\rm mol}$ the
  molecular ratio (the relative attractiveness of the molecule relative
  to the $P_{\psi}^N(4312)$) as defined in Eq.~(\ref{eq:Rmol}),
  $M_{\rm mol}$ the mass of the molecule,   ``Candidate'' refers to
  an experimentally observed pentaquark that could
  correspond with our theoretical prediction and
  $M_{\rm candidate}$ to its mass.
  Masses are expressed in units of ${\rm MeV}$.
  In the $M_{\rm mol}$ column the superscript ``$V$'' indicates a virtual state.
  The errors in $M_{\rm mol}$ come from propagating the uncertainties in the mass
  of the input or reference state ($P_{\psi}^N(4312)$), the mass of the scalar
  meson, \corr{the cutoff and the choice of the parameter $\alpha$} and
  then choosing the largest of them.
  When the uncertainties allow a state to change from virtual to bound or
  vice versa, we indicate it by adding $(V)$ or $(B)$ after
  the upper error.
}
\label{tab:predictions-triplet-antitriplet}
\end{table}

\begin{table}[!ttt]
\begin{tabular}{|clllcl|}
\hline\hline
System  & $I$($J^{P}$) & $R_{\rm mol}$ & $M_{\rm mol}$ &
Candidate & $M_{\rm candidate}$\\
\hline
  $\Sigma_c \bar{D}$ & $\tfrac{1}{2}$ ($\tfrac{1}{2}^{-}$) & $1.00$ 
& Input & $P_{\psi}^N(4312)$ & $4311.9^{+6.8}_{-0.9}$ \\
  $\Sigma_c^* \bar{D}$ & $\tfrac{1}{2}$ ($\tfrac{3}{2}^{-}$) & \corr{$1.01$} 
  & $4376.0^{+7.1}_{-0.9}$ & - & - \\
$\Sigma_c \bar{D}^*$ & $\tfrac{1}{2}$ ($\tfrac{1}{2}^{-}$) & $0.85$
& $4459.7^{+2.3}_{-2.5}$ & $P_{\psi}^N(4457)$ & $4457.3^{+4.1}_{-1.8}$ \\
$\Sigma_c \bar{D}^*$ & $\tfrac{1}{2}$ ($\tfrac{3}{2}^{-}$) & $1.13$
& \corr{$4445.2^{+10.4}_{-4.7}$} & $P_{\psi}^N(4440)$  & $4440.3^{+4.3}_{-1.8}$ \\
  $\Sigma_c^* \bar{D}^*$ & $\tfrac{1}{2}$ ($\tfrac{1}{2}^{-}$) & $0.82$  
 & $4525.4^{+1.3(V)}_{-2.7}$ & -  & - \\
$\Sigma_c^* \bar{D}^*$ & $\tfrac{1}{2}$ ($\tfrac{3}{2}^{-}$) & $0.96$
& $4520.3^{+5.3}_{-1.7}$ & -  & - \\
$\Sigma_c^* \bar{D}^*$ & $\tfrac{1}{2}$ ($\tfrac{5}{2}^{-}$) & $1.19$
& \corr{$4505.8^{+12.0}_{-7.5}$} & -  & - \\
\hline\hline 
$\Xi_c' \bar{D}$ & $0$ ($\tfrac{1}{2}^{-}$) & $1.02$ 
& $4435.9^{+7.3}_{-0.9}$ & - & - \\
$\Xi_c^* \bar{D}$ & $0$ ($\tfrac{3}{2}^{-}$) & $1.03$ 
& $4502.5^{+7.5}_{-1.0}$ & - & - \\
$\Xi_c' \bar{D}^*$ & $0$ ($\tfrac{1}{2}^{-}$) & $0.88$ 
& $4584.2^{+2.8}_{-2.7}$ & - & - \\
$\Xi_c' \bar{D}^*$ & $0$ ($\tfrac{3}{2}^{-}$) & $1.16$ 
& \corr{$4568.8^{+10.3}_{-5.8}$} & - & - \\
$\Xi_c^* \bar{D}^*$ & $0$ ($\tfrac{1}{2}^{-}$) & $0.84$ 
& $4652.5^{+1.7(V)}_{-3.0}$ & - & - \\
$\Xi_c^* \bar{D}^*$ & $0$ ($\tfrac{3}{2}^{-}$) & $0.98$ 
& $4646.9^{+5.8}_{-1.8}$ & - & - \\
$\Xi_c^* \bar{D}^*$ & $0$ ($\tfrac{5}{2}^{-}$) & $1.22$ 
& \corr{$4631.8^{+12.5}_{-8.7}$} & - & - \\
\hline\hline
$\Sigma_c \bar{D}_s$-$\Xi_c' \bar{D}$ & $1$ ($\tfrac{1}{2}^{-}$) & $0.96$
& $4417.3^{+4.2}_{-1.6}$ & - & - \\
$\Sigma_c^* \bar{D}_s$-$\Xi_c^* \bar{D}$ & $1$ ($\tfrac{3}{2}^{-}$) & $0.98$
& $4481.6^{+4.4}_{-1.7}$ & - & - \\
$\Sigma_c \bar{D}_s^*$-$\Xi_c' \bar{D}^*$ & $1$ ($\tfrac{1}{2}^{-}$) & $0.84$ 
& \corr{$4581.9^{+4.4}_{-2.4} - (0.6^{+0.6}_{-0.5})\, i$} & - & - \\
$\Sigma_c \bar{D}_s^*$-$\Xi_c' \bar{D}^*$ & $1$ ($\tfrac{3}{2}^{-}$) & $1.08$ 
& \corr{$4556.4^{+7.4}_{-2.6}$} & - & - \\
$\Sigma_c^* \bar{D}_s^*$-$\Xi_c^* \bar{D}^*$ & $1$ ($\tfrac{1}{2}^{-}$) & $0.81$
& \corr{$4647.4^{+5.3}_{-2.5} - (1.8^{+2.7}_{-1.2})\, i$} & - & - \\
$\Sigma_c^* \bar{D}_s^*$-$\Xi_c^* \bar{D}^*$ & $1$ ($\tfrac{3}{2}^{-}$) & $0.94$ 
& $4625.5^{+4.3}_{-2.6}$ & - & - \\
$\Sigma_c^* \bar{D}_s^*$-$\Xi_c^* \bar{D}^*$ & $1$ ($\tfrac{5}{2}^{-}$) & $1.13$ 
& \corr{$4618.6^{+8.7}_{-5.1}$} & - & - \\
\hline\hline
$\Xi_c' \bar{D}_s$-$\Omega_c \bar{D}$ & $\tfrac{1}{2}$ ($\tfrac{1}{2}^{-}$) & $1.01$ 
& $4542.9^{+3.9}_{-1.8}$ & - & - \\
$\Xi_c^* \bar{D}_s$-$\Omega_c^* \bar{D}$ & $\tfrac{1}{2}$ ($\tfrac{3}{2}^{-}$) & $1.02$
& $4610.0^{+3.9}_{-2.0}$ & - & - \\
$\Xi_c' \bar{D}_s^*$-$\Omega_c \bar{D}^*$ & $\tfrac{1}{2}$ ($\tfrac{1}{2}^{-}$) & $0.87$ 
& \corr{$4699.6^{+3.8}_{-2.9} - (0.5^{+0.4}_{-0.3})\, i$} & - & - \\
$\Xi_c' \bar{D}_s^*$-$\Omega_c \bar{D}^*$ & $\tfrac{1}{2}$ ($\tfrac{3}{2}^{-}$) & $1.13$ 
& \corr{$4681.0^{+8.3}_{-5.1}$} & - & - \\
$\Xi_c^* \bar{D}_s$-$\Omega_c^* \bar{D}$ & $\tfrac{1}{2}$ ($\tfrac{1}{2}^{-}$) & $0.84$ 
& \corr{$4770.8^{+4.2}_{-3.5} - (1.0^{+0.3}_{-0.6})\, i$} & - & - \\
$\Xi_c^* \bar{D}_s$-$\Omega_c^* \bar{D}$ & $\tfrac{1}{2}$ ($\tfrac{3}{2}^{-}$) & $0.97$ 
& $4752.0^{+4.9}_{-2.7}$ & - & - \\
$\Xi_c^* \bar{D}_s$-$\Omega_c^* \bar{D}$ & $\tfrac{1}{2}$ ($\tfrac{5}{2}^{-}$) & $1.19$ 
& \corr{$4746.3^{+9.6}_{-8.1}$} & - & - \\
\hline\hline
\end{tabular}
\caption{Molecular octet pentaquark states predicted in this work.
  We refer to Table \ref{tab:predictions-triplet-antitriplet}
  for the conventions we follow.
  A complex value of $M_{\rm mol}$ indicates that the state is a resonance,
  i.e. a pole in the $(II,I)$ Riemann sheet with respect to the
  lower and higher mass threshold.
  If there is a $V$ superscript above the complex $M_{\rm mol}$, this indicates 
  a pole in the $(I,II)$ sheet instead.
  All binding energies and masses are in units of ${\rm MeV}$.
}
\label{tab:predictions-octet}
\end{table}

\begin{table}[ttt]
\begin{tabular}{|clll|}
\hline\hline
System  & $I$($J^{P}$) & $R_{\rm mol}$ & $M_{\rm mol}$ \\
\hline
  $\Sigma_c \bar{D}$ & $\tfrac{3}{2}$ ($\tfrac{1}{2}^{-}$) & $0.57$
  & \corr{$(4316.8^{+3.9(B)}_{-27.3})^V$} \\
  $\Sigma_c^* \bar{D}$ & $\tfrac{3}{2}$ ($\tfrac{3}{2}^{-}$) & $0.58$ 
  & \corr{$(4381.8^{+3.6(B)}_{-26.4})^V$} \\
  $\Sigma_c \bar{D}^*$ & $\tfrac{3}{2}$ ($\tfrac{1}{2}^{-}$) & $0.96$ 
& $4455.6^{+5.4}_{-1.5}$ \\
$\Sigma_c \bar{D}^*$ & $\tfrac{3}{2}$ ($\tfrac{3}{2}^{-}$) & $0.41$  & - \\
  $\Sigma_c^* \bar{D}^*$ & $\tfrac{3}{2}$ ($\tfrac{1}{2}^{-}$) & $1.07$ 
 & \corr{$4514.6^{+8.3}_{-1.9}$} \\
$\Sigma_c^* \bar{D}^*$ & $\tfrac{3}{2}$ ($\tfrac{3}{2}^{-}$) & $0.79$
& $4526.1^{+0.5(V)}_{-2.6}$  \\
$\Sigma_c^* \bar{D}^*$ & $\tfrac{3}{2}$ ($\tfrac{5}{2}^{-}$) &
$0.32$  & - \\
\hline\hline
$\Sigma_c \bar{D}_s$-$\Xi_c' \bar{D}$ & $1$ ($\tfrac{1}{2}^{-}$) & $0.65$ 
& \corr{$(4445.8^{+0.0}_{-2.2(R)} - (0.1^{+7.1}_{-0.1(R)})\,i)^V$}  \\
$\Sigma_c^* \bar{D}_s$-$\Xi_c^* \bar{D}$ & $1$ ($\tfrac{3}{2}^{-}$) & $0.65$
& \corr{$(4513.0^{+0.0}_{-2.2(R)} - (0.0^{+6.7}_{-0.0(R)})\,i)^V$}  \\
$\Sigma_c \bar{D}_s^*$-$\Xi_c' \bar{D}^*$ & $1$ ($\tfrac{1}{2}^{-}$) & $1.02$ 
& $4560.2^{+4.8}_{-2.0}$ \\
$\Sigma_c \bar{D}_s^*$-$\Xi_c' \bar{D}^*$ & $1$ ($\tfrac{3}{2}^{-}$) & $0.50$
& \corr{$(4587.5^{+13.5}_{-2.4(R)} - (4.6^{+22.9}_{-4.6(R)})\, i)^V$} \\
$\Sigma_c^* \bar{D}_s^*$-$\Xi_c^* \bar{D}^*$ & $1$ ($\tfrac{1}{2}^{-}$)
& $1.12$ 
& \corr{$4622.5^{+6.6}_{-2.0}$}  \\
$\Sigma_c^* \bar{D}_s^*$-$\Xi_c^* \bar{D}^*$ & $1$ ($\tfrac{3}{2}^{-}$) & $0.85$
& $4651.9^{+2.3}_{-3.0} - (0.0\pm 0.0)\, i$ \\
$\Sigma_c^* \bar{D}_s^*$-$\Xi_c^* \bar{D}^*$ & $1$ ($\tfrac{5}{2}^{-}$) & $0.42$ 
& \corr{$(4656.2^{+24.3}_{-3.0(R)} - (9.8^{+31.6}_{-9.8(R)}) \,i)^V$} \\
\hline\hline
$\Xi_c' \bar{D}_s$-$\Omega_c \bar{D}$ & $\tfrac{1}{2}$ ($\tfrac{1}{2}^{-}$) &
$0.72$
& \corr{$4560.2^{+4.6}_{-2.9} - (1.5^{+1.0}_{-1.2})\, i$}  \\
$\Xi_c^* \bar{D}_s$-$\Omega_c^* \bar{D}$ & $\tfrac{1}{2}$ ($\tfrac{3}{2}^{-}$) & $0.73$
 & \corr{$4630.6^{+4.3}_{-2.8} - (1.5^{+1.0}_{-1.2})\, i$}  \\
$\Xi_c' \bar{D}_s^*$-$\Omega_c \bar{D}^*$ & $\tfrac{1}{2}$ ($\tfrac{1}{2}^{-}$) & $1.07$ 
& \corr{$4680.3^{+7.6}_{-1.8}$}  \\
$\Xi_c' \bar{D}_s^*$-$\Omega_c \bar{D}^*$ & $\tfrac{1}{2}$ ($\tfrac{3}{2}^{-}$) & $0.59$ 
& \corr{$4704.9^{+18.1}_{-6.2} - (4.7^{+5.3}_{-2.6})\, i$}  \\
$\Xi_c^* \bar{D}_s$-$\Omega_c^* \bar{D}$ & $\tfrac{1}{2}$ ($\tfrac{1}{2}^{-}$) & $1.17$ 
& \corr{$4742.8^{+9.9}_{-5.3}$} \\
$\Xi_c^* \bar{D}_s$-$\Omega_c^* \bar{D}$ & $\tfrac{1}{2}$ ($\tfrac{3}{2}^{-}$) & $0.92$ 
& $4769.3^{+4.5}_{-2.7} - (0.1 \pm 0.0)\, i$ \\
$\Xi_c^* \bar{D}_s$-$\Omega_c^* \bar{D}$ & $\tfrac{1}{2}$ ($\tfrac{5}{2}^{-}$) & $0.52$ 
& \corr{$4778.2^{+27.3}_{-8.5} - (6.5^{+8.6}_{-3.5})\, i$} \\
\hline\hline
  $\Omega_c \bar{D}_s$ & $0$ ($\tfrac{1}{2}^{-}$) & $0.78$ 
& $4663.1^{+0.5(V)}_{-2.8}$ \\
  $\Omega_c^* \bar{D}_s$ & $0$ ($\tfrac{3}{2}^{-}$) & $0.79$ 
  & $4733.6^{+0.7(V)}_{-3.0}$ \\
  $\Omega_c \bar{D}_s^*$ & $0$ ($\tfrac{1}{2}^{-}$) & $1.12$ 
& \corr{$4792.9^{+9.2}_{-3.7}$} \\
$\Omega_c \bar{D}_s^*$ & $0$ ($\tfrac{3}{2}^{-}$) & $0.67$ 
& \corr{$(4807.0^{+0.4(B)}_{-13.6})^V$} \\
  $\Omega_c^* \bar{D}_s^*$ & $0$ ($\tfrac{1}{2}^{-}$) & $1.20$ 
 & \corr{$4857.9^{+11.5}_{-7.6}$} \\
$\Omega_c^* \bar{D}_s^*$ & $0$ ($\tfrac{3}{2}^{-}$) & $0.98$ 
& $4871.4^{+5.4}_{-2.8}$ \\
  $\Omega_c^* \bar{D}_s^*$ & $0$ ($\tfrac{5}{2}^{-}$) & $0.60$ 
& \corr{$(4875.9^{+2.2(B)}_{-30.0})^V$} \\
\hline \hline
\end{tabular}
\caption{Molecular decuplet pentaquark states predicted in this work.
  We refer to Tables \ref{tab:predictions-triplet-antitriplet}
  and \ref{tab:predictions-octet} for conventions.
  However no experimental candidate is known for the decuplet
  and thus we have not included the columns regarding candidates.
  If the uncertainties in the mass of a pole originally located in the $(I,II)$
  Riemann sheet are able to move it to the $(II,I)$ sheet (i.e. the sheet
  corresponding to resonances), we will indicate this with the letter $(R)$
  following the lower error.
  All binding energies and masses are in units of ${\rm MeV}$.
}
\label{tab:predictions-decuplet}
\end{table}

In a first approximation
we will ignore all the coupled channel dynamics except for the $S = -1$,
$I = 1$ sector of $\bar{H}_c S_c$, i.e. $\bar{D}_s \Sigma_c$-$\bar{D} \Xi_c^{'}$
and its analogues involving excited charmed antimesons and baryons.
That is, we will only consider what we called type (a) coupled channel dynamics
in the previous subsection.
The coupled channel potential for this sector takes the generic form
\begin{eqnarray}
  V(\bar{D}_s \Sigma_c-\bar{D} \Xi_c^{'}) =
  \begin{pmatrix}
    \frac{2}{3} V^{O} + \frac{1}{3} V^{D} & - \frac{\sqrt{2}}{3}(V^O - V^D) \\
    - \frac{\sqrt{2}}{3}(V^O - V^D) & \frac{1}{3} V^{O} + \frac{2}{3} V^{D}
  \end{pmatrix} \, , \nonumber \\ \label{eq:V-CC-1} \\
    V(\bar{D}_s \Xi_c^{'}-\bar{D} \Omega_c) =
  \begin{pmatrix}
    \frac{1}{3} V^{O} + \frac{2}{3} V^{D} & - \frac{\sqrt{2}}{3}(V^O - V^D) \\
    - \frac{\sqrt{2}}{3}(V^O - V^D) & \frac{2}{3} V^{O} + \frac{1}{3} V^{D}
  \end{pmatrix} \, , \nonumber \\ \label{eq:V-CC-2}
\end{eqnarray}
where $V^O$ and $V^D$ refer to the octet and decuplet potential.
In terms of the saturated contact-range couplings, and ignoring in a first
approximation the difference among the vector meson masses ($m_{\rho}$,
$m_{\omega}$, $m_{K^*}$, $m_{\phi}$), we have
\begin{eqnarray}
  V^O  = C^O &\propto& 
  -\frac{g_{V1} g_{V2}}{m_V^2}
  \left( 1 + \kappa_{V1} \kappa_{V2} \, \frac{m_V^2}{6 M^2}\,\hat{C}_{L12} \right)
  \nonumber \\
  && \quad \,\, - {(\frac{m_V}{m_S})}^{\alpha}\frac{g_{S1} g_{S2}}{m_S^2} \, , \\
  V^D = C^D &\propto& 
  +2\,\frac{g_{V1} g_{V2}}{m_V^2}
  \left( 1 + \kappa_{V1} \kappa_{V2} \, \frac{m_V^2}{6 M^2}\,\hat{C}_{L12} \right)
  \nonumber \\
  && \quad \,\, - {(\frac{m_V}{m_S})}^{\alpha}\frac{g_{S1} g_{S2}}{m_S^2} \, ,
\end{eqnarray}
where the subindices $1$ and $2$ indicate the charmed antimeson and charmed
sextet baryon, respectively.
That is, the octet and decuplet potentials correspond to taking
the G-parity and isospin factors $[ \zeta + T_{12}] = -1$ and $2$
in Eqs.~(\ref{eq:V_S}), (\ref{eq:V_V}) and (\ref{eq:coupling-sat}),
to which we refer for further details.
Besides this, when the exchanged vector meson is not the $\rho$ or
the $\omega$, we have to correct the previous expressions
to take into account the heavier $K^*$ and $\phi$ masses.
To provide a concrete example, we might consider the non-diagonal components
of the coupled channel potentials of Eqs.~(\ref{eq:V-CC-1}) and
(\ref{eq:V-CC-2}), which require the exchange of a $K^*$ meson,
leading to
\begin{eqnarray}
  && V(\bar{D}_s \Sigma_c \to \bar{D} \Xi_c(1)) =
  V(\bar{D}_s \Xi_c \to \bar{D} \Omega_c) \nonumber \\
  && \qquad = - \frac{\sqrt{2}}{3}(V^O - V^D) 
  \nonumber \\
  &&  \qquad \propto {(\frac{m_V}{m_{K^*}})}^{\alpha} \frac{g_{V1} g_{V2}}{m_{K^*}^2}
  \left( 1 + \kappa_{V1} \kappa_{V2} \, \frac{m_{K^*}^2}{6 M^2}\,\hat{C}_{L12} \right) \, , 
\end{eqnarray}
where we remind that $m_V = (m_{\rho} + m_{\omega})/2$. 

If we use the $P_{\psi}^N(4312)$ as the reference state, we obtain
the predictions contained in Tables \ref{tab:predictions-triplet-antitriplet},
\ref{tab:predictions-octet} and \ref{tab:predictions-decuplet}.
The first set of predictions --- Table \ref{tab:predictions-triplet-antitriplet}
--- refers to the $\bar{H}_c T_c$ type of pentaquarks.
This includes the $P_{\psi s}^{\Lambda}(4338)$ and $P_{\psi s}^{\Lambda}(4459)$,
which we predict to be $I=0$ $\bar{D} \Xi_c$ and $\bar{D}^* \Xi_c$
bound states with masses of $4327$ and $4467\,{\rm MeV}$,
both within $10\,{\rm MeV}$ of their
experimental masses.
The two $\bar{D}^* \Xi_c$ states (with total spin $J=\tfrac{1}{2}$ and
$\tfrac{3}{2}$, respectively) are degenerate within our choice of
coupled channel dynamics.
Only when the nearby $\bar{D} \Xi_c'$ and $\bar{D} \Xi_c^*$ channels are
included do we obtain a hyperfine splitting between the two spin
configurations~\cite{Peng:2020hql,Yan:2021nio} (we include these channels
in the next subsection, where results are shown
Table \ref{tab:predictions-coupled-channels}),
hence reproducing the two peak solution
in~\cite{Aaij:2020gdg}.
It is also worth noticing the prediction of an $I=1$ $\bar{D} \Xi_c$ molecule
with a mass of $4336\,{\rm MeV}$, really close to the experimental
mass of the $P_{\psi s}^{\Lambda}(4338)$, and which might indicate
the possible importance of isospin breaking effects and
mixing of near threshold $P_{\psi s}^{\Lambda}$ and
$P_{\psi s}^{\Sigma}$ pentaquarks~\cite{Yan:2022wuz,Meng:2022wgl}.
There is also a $\bar{D}_s \Lambda_c$ bound state with a mass of
$4252\,{\rm MeV}$, which is similar to the $P_{\psi s}^{\Lambda}(4255)$
state predicted in~\cite{Yan:2022wuz} as a flavor partner of
the $P_{\psi s}^{\Lambda}(4338)$ (or in~\cite{Yan:2021nio}
as a flavor and HQSS partner of the $P_{\psi s}^{\Lambda}(4459)$).
Recently, the amplitude analysis of the $J/\psi \Lambda$ invariant mass
distribution of~\cite{Nakamura:2022jpd} also found a $P_{\psi s}^{\Lambda}(4255)$
pentaquark as a $\bar{D}_s \Lambda_c$ virtual state
extremely close to threshold.
Meanwhile in~\cite{Chen:2022wkh}, which proposes a different phenomenological
contact-range potential where the couplings are directly saturated
from the light-quark interaction, the existence of a bound
$P_{\psi s}^{\Lambda}(4255)$ is neither confirmed or denied
but remains contingent on the choice of couplings
(though~\cite{Chen:2022wkh} does not look
for virtual state solutions).

Table \ref{tab:predictions-triplet-antitriplet} also predicts three
nucleon-like $P^N_{\psi}$ states close to threshold:
a $P^N_{\psi} (4154)$, and two $P^N_{\psi}(4295)$ with spins
$J = \tfrac{1}{2}$, $\tfrac{3}{2}$ and degenerate masses.
The $P^N_{\psi}(4154)$ might correspond with a possible structure in the
$J/\psi \bar{p}$ invariant mass briefly mentioned in~\cite{LHCb:2022jad},
though there is a statistical preference for a model
without this resonant contribution.
Very recently, the new GlueX data~\cite{Adhikari:2023fcr}
for $\gamma p \to J/\psi p$ shows a dip between the $\bar{D} \Lambda_c$ and
$\bar{D}^* \Lambda_c$ thresholds, which in~\cite{Strakovsky:2023kqu}
has been interpreted as compatible with a pentaquark signal.
The fit in~\cite{Strakovsky:2023kqu} results in a mass of
$M = 4235 \pm 8\,{\rm MeV}$, i.e.,
in between the $P^N_{\psi} (4154)$ and $P^N_{\psi}(4295)$ masses.
Yet, the analysis of~\cite{Strakovsky:2023kqu}
is focused on the compatibility between the GlueX data~\cite{Adhikari:2023fcr}
and the existence of the $P^N_{\psi}$ pentaquarks already observed
by the LHCb, and not on finding new $P^N_{\psi}$ pentaquarks.

The second set of predictions --- Table \ref{tab:predictions-octet} ---
pertains the octet $\bar{H}_c S_c$ pentaquarks.
Besides the reference state, the $P_{\psi}^N(4312)$, other two candidate states
are known, the $P_{\psi}^N(4440)$ and $P_{\psi}^N(4457)$, which we reproduce as
the $J=\tfrac{3}{2}$ and $J=\tfrac{1}{2}$ $\bar{D}^* \Sigma_c$ bound states,
respectively.
Here it is worth commenting that the saturation model provides a very
specific prediction of the spin of the $P_{\psi}^N(4440)$ and
$P_{\psi}^N(4457)$ as molecules,
which agrees with other phenomenological molecular models~\cite{Liu:2019zvb,Yamaguchi:2019seo}
(but not all of them~\cite{Wang:2019nvm}).
In contrast compact pentaquark models usually predict the opposite spin
identification~\cite{Garcilazo:2022kra},
with the $P_{\psi}^N(4440)$ and $P_{\psi}^N(4457)$ being
$J = \tfrac{1}{2}$ and $J=\tfrac{3}{2}$, respectively.
Experimentally, the spins of these two pentaquarks have not been determined yet.
There is a fourth $P^N_{\psi}$ pentaquark candidate that is consistently
predicted in most molecular models~\cite{Liu:2019tjn,Liu:2019zvb,Guo:2019fdo,Pan:2019skd}
(as a consequence of HQSS) and
that a later theoretical analysis of the LHCb $J/\psi p$ data has
shown to exist~\cite{Du:2019pij}, the $P_{\psi}^N(4380)$.
Unsurprisingly (as we incorporate HQSS),
our model also reproduces this state.
The predictions we obtain for the $P_{\psi}^N$ pentaquarks are also similar
to {\it scenario B}~\footnote{Scenario A and B refer to the spin of
  the $P^N_{\psi}(4440)$ and $P^N_{\psi}(4457)$ in their interpretation
  as $\bar{D}^* \Sigma_c$ bound states: A is $J=\tfrac{1}{2}$ and
  $\tfrac{3}{2}$, respectively, while B is the opposite
  identification. Scenario B is equivalent to
  the spin prediction in our model.}
in Refs.~\cite{Liu:2019tjn,Valderrama:2019chc},
where a pionless and pionful EFT were used to deduce
the $\bar{D}^{(*)} \Sigma_c^{(*)}$ molecular spectrum.

{
  It should be noted that we previously included OPE explicitly for the
  $\Sigma_c \bar{D}^*$ and $\Sigma_c^* \bar{D}^*$ molecules
  in Ref.~\cite{Peng:2021hkr} (whose results are
  basically equivalent to the ones presented here~\footnote{ 
  It is the same model with the same input, the only difference being
  tiny changes in the charmed meson masses, as Ref.~\cite{Peng:2021hkr}
  uses a previous version of the RPP. These changes only manifest
  as truncation errors in the last digit (i.e. as occasional
  discrepancies of $0.1\,{\rm MeV}$ in the masses of
  the pentaquarks).}).
  There we observed that OPE shifts the masses of these pentaquarks
  by about $(1-2)\,{\rm MeV}$; once its sign is taken
  into account, we end up with
  \begin{eqnarray}
    M_{\rm OPE}(\Sigma_c \bar{D}^*, J=\tfrac{1}{2}) &=&
    4461.3\,(4459.7^{+2.3}_{-2.5})\,{\rm MeV} \, , \\
    M_{\rm OPE}(\Sigma_c \bar{D}^*, J=\tfrac{3}{2}) &=&
    4443.6\,(4445.2^{+10.4}_{-4.7})\,{\rm MeV} \, , \\
    \nonumber \\
    M_{\rm OPE}(\Sigma_c^* \bar{D}^*, J=\tfrac{1}{2}) &=&
    4526.7\,(4525.4^{+1.3(V)}_{-2.7})\,{\rm MeV} \, , \\
    M_{\rm OPE}(\Sigma_c^* \bar{D}^*, J=\tfrac{3}{2}) &=&
    4521.5\,(4520.3^{+5.3}_{-1.7})\,{\rm MeV} \, , \\
    M_{\rm OPE}(\Sigma_c^* \bar{D}^*, J=\tfrac{5}{2}) &=&
    4503.3\,(4505.8^{+12.0}_{-7.5})\,{\rm MeV} \, , 
  \end{eqnarray}
  where the number in parentheses is the original calculation
  without OPE from Table \ref{tab:predictions-octet}.
  It happens that the correction from OPE falls in all cases
  within the uncertainties of our calculation.
  For other pentaquarks the mass shifts from OPE are expected to be smaller:
  in all the other configurations from Tables \ref{tab:predictions-octet}
  and \ref{tab:predictions-decuplet} OPE is not as strong
  as in the $I=1/2$ $\Sigma_c \bar{D}^*$ and
  $\Sigma_c^* \bar{D}^*$ systems.
  For further details on the calculations with OPE, we refer to Appendix C
  within~\cite{Peng:2021hkr}.
}

In addition,
Table \ref{tab:predictions-octet} contains the predictions of a series of
$P_{\psi s}^{\Lambda}$, $P_{\psi s}^{\Sigma}$ and $P_{\psi ss}^{\Xi}$ pentaquarks.
There are previous molecular predictions of $P_{\psi s}^{\Lambda}$ pentaquarks~\cite{Xiao:2019gjd} and of $P_{\psi s}^{\Sigma}$ and $P_{\psi ss}^{\Xi}$
states~\cite{Peng:2019wys} that are in line
with those we obtain here.
In particular the comparison with the EFT of Ref.~\cite{Peng:2019wys} is
interesting, as most of our predictions are very similar to scenario B
in~\cite{Peng:2019wys}.
The only exceptions are the $J=\tfrac{1}{2}$
$\bar{D}_s^* \Sigma_c$-$\bar{D}^* \Xi_c'$
($\bar{D}_s^* \Xi_c'$-$\bar{D}^* \Omega_c$) and
$\bar{D}_s^* \Sigma_c^*$-$\bar{D}^* \Xi_c^*$
($\bar{D}_s^* \Xi_c'$-$\bar{D}^* \Omega_c$)
pentaquark states (i.e. the four resonances
in Table \ref{tab:predictions-octet}), for which the decuplet
configurations are the most attractive (thus in contradiction
with the original assumption made in~\cite{Peng:2019wys}
that the octet was the most attractive configuration,
hence the difference).
We notice that the idea of a pentaquark octet was previously conjectured
in~\cite{Santopinto:2016pkp}, which assumes compact,
non-molecular pentaquarks, and it has also appeared
in the hadro-charmonium picture~\cite{Eides:2017xnt,Ferretti:2020ewe}.
Here it is worth noticing that the predictions of molecular and non-molecular
pentaquarks differ in their masses: the mass difference among $P_{\psi}^N$,
$P_{\psi s}^{\Lambda}$, $P_{\psi s}^{\Sigma}$ and $P_{\psi ss}^{\Xi}$
pentaquarks depends on their nature, i.e. if these pentaquarks are observed
spectroscopy alone would be able to provide a clue about whether they are
molecular or not.

The third set of predictions, shown in Table \ref{tab:predictions-decuplet},
is for the decuplet $\bar{H}_c S_c$ pentaquarks, of which
no experimental candidate is known.
It is interesting to notice the prediction of a $P_{\psi}^{\Delta}$ molecular
pentaquark with a mass of $4456\,{\rm MeV}$.
This is really close to the $\bar{D}^{*0} \Sigma_c^+$ and ${D}^{*-} \Sigma_c^{++}$
thresholds, just like the experimental $P_{\psi}^N(4457)$, thus suggesting
the possibility of sizable isospin breaking effects,
as investigated in~\cite{Guo:2019fdo}.
Previous predictions of $P_{\psi}^{\Delta}$ or other decuplet pentaquarks are
scarce in the molecular case, owing to the fact that the electric-type
component of vector meson exchange is repulsive,
though they appear in the one boson exchange models
of~\cite{Chen:2015loa,Liu:2019zvb,Wang:2021hql} (which usually
include scalar meson exchange or the tensor components of
vector meson exchange).

For the octet and decuplet predictions of Tables \ref{tab:predictions-octet}
and \ref{tab:predictions-decuplet}, there is the problem of
how to distinguish the octet or decuplet nature of a state
when there are coupled channel dynamics.
This is done by considering the vertex functions $\phi_A$ (with $A = 1,2$)
for the $\bar{D} \Xi_c'$-$\bar{D}_s \Sigma_c$ and
$\bar{D}_s \Xi_c'$-$\bar{D} \Omega_c$ family of
pentaquarks.
For the bound state solutions the relative signs of the vertex functions for
the two channels indicate whether we are dealing with an octet or a decuplet
(by inverting Eqs.~(\ref{eq:HS-decomp-1}-\ref{eq:HS-decomp-4})).
This is not the case though when the pole is above the lower threshold, i.e.
when we have a resonance, as the vertex functions become complex numbers.
Yet, it happens that for every solution above the lower threshold we have found
a second solution below it, from which it is trivial to determine
the SU(3)-flavor representation of the resonant solution.
To give an example, in the $\bar{D}_s \Xi_c'$-$\bar{D} \Omega_c$ system
we find two poles with masses and vertex functions:
\begin{eqnarray}
  M_L &=& 4542.9\,{\rm MeV} \, , \\
  {(\phi_1,\phi_2)}_L &=& (0.84,-0.54) \, , \\
  \nonumber \\
  M_H &=& (4560.2 - 1.5\,i)\,{\rm MeV}\, ,  \\
  {(\phi_1,\phi_2)}_H &=& (0.01-0.31\,i, 0.95) \, ,
\end{eqnarray}
for the lower and higher mass poles (subscripts $L$ and $H$), respectively.
For this system a pure octet configuration corresponds to
$(\phi_1, \phi_2) = (\sqrt{2/3},-\sqrt{1/3}) \simeq (0.82,-0.58)$.
From this, it is apparent that the vertex functions of the lower mass pole 
indicate a predominantly octet state, which we include
in Table \ref{tab:predictions-octet}.
Then, by our convention, the higher mass state is considered to be a decuplet
and we include it in Table \ref{tab:predictions-decuplet}.

\subsection{Predictions with extended coupled channel dynamics}

\begin{table*}[!ttt]
\begin{tabular}{|ccccllcl|}
\hline\hline
Channels & System & $I$($J^{P}$) & $S$  & $M^{\rm CC}_{\rm mol}$
& $\Delta M^{\rm CC}_{\rm mol}$ & Candidate & $M_{\rm candidate}$\\
\hline
\multirow{2}{*}{(1) $\Lambda_c \bar{D}^*$-$\Sigma_c \bar{D}$}
& $\Lambda_c \bar{D}^*$ & $0$($\frac{1}{2}^-$) & $0$ & \corr{$4291.0^{+3.9}_{-2.4}$} & $-4.0^{(*)}$ & - & - \\
& $\Sigma_c \bar{D}$ & $0$($\frac{1}{2}^-$) & $0$ &
\corr{$4315.8^{+5.6}_{-2.3} - (6.6^{+9.6}_{-4.2})\,i$} & $+3.9$ &
$P^N_{\psi}(4312)$ & $4311.9^{+6.8}_{-0.9} - (4.9^{+2.3}_{-2.6})\,i$ \\
\hline\hline
\multirow{2}{*}{(2) $\Lambda_c \bar{D}_s$-$\Xi_c \bar{D}$}
& $\Lambda_c \bar{D}_s$ & $0$($\frac{1}{2}^-$) & $-1$ &
$4251.8^{+3.0}_{-1.6}$ & $-0.6$ & - & - \\
& $\Xi_c \bar{D}$ & $0$($\frac{1}{2}^-$) & $-1$ &
\corr{$4328.1^{+6.4}_{-0.8} - (1.1^{+2.2}_{-0.7})\,i$} & $+0.7$ &
$P^{\Lambda}_{\psi s}(4338)$ & $4338.2 \pm 0.7 - (3.5 \pm 0.6)\,i$ \\
\hline \hline
\multirow{3}{*}{(3) $\Lambda_c \bar{D}_s^*$-$\Xi_c' \bar{D}$-$\Xi_c \bar{D}^*$}
& $\Lambda_c \bar{D}_s^*$ & $0$($\frac{1}{2}^-$) & $-1$ & $4393.5^{+4.7}_{-1.3}$ & $-1.7$ & - & - \\
& $\Xi_c' \bar{D}$ & $0$($\frac{1}{2}^-$) & $-1$ &
\corr{$4433.6^{+8.8}_{-2.8} - (0.5^{+0.4}_{-0.3})\,i$} & $-4.3$ & - & - \\
& $\Xi_c \bar{D}^*$ & $0$($\frac{1}{2}^-$) & $-1$ & \corr{$4468.8^{+6.4}_{-1.3} - (3.0^{+5.2}_{-2.0})\,i$} & $+1.1$ & $P^{\Lambda}_{\psi s}(4459)$ & $4458.9^{+5.5}_{-3.1} - (8.7^{+5.2}_{-4.3})\,i$ \\
\hline \hline
\multirow{3}{*}{(4) $\Lambda_c \bar{D}_s^*$-$\Xi_c \bar{D}^*$-$\Xi_c^* \bar{D}$}
& $\Lambda_c \bar{D}_s^*$ & $0$($\frac{3}{2}^-$) & $-1$ & $4394.0^{+4.3}_{-1.6}$ & $-1.2$ & - & - \\
& $\Xi_c \bar{D}^*$ & $0$($\frac{3}{2}^-$) & $-1$ & \corr{$4464.6^{+9.1}_{-4.3} - (0.6^{+0.7}_{-0.4})\,i$} & $-2.1$ & $P^{\Lambda}_{\psi s}(4459)$ & $4458.9^{+5.5}_{-3.1} - (8.7^{+5.2}_{-4.3})\,i$ \\
& $\Xi_c' \bar{D}$ & $0$($\frac{3}{2}^-$) & $-1$ & \corr{$4503.8^{+6.7}_{-0.9} - (2.0^{+3.5}_{-1.3})\,i$} & $+1.3$ &  - & - \\
\hline \hline
\multirow{4}{*}{(5) $\Sigma_c \bar{D}_s$-$\Xi_c' \bar{D}$-$\Xi_c \bar{D}^*$}
& $\Sigma_c \bar{D}_s$-$\Xi_c' \bar{D}$ (O)
& $1$($\frac{1}{2}^-$) & $-1$ & \corr{$4415.2^{+5.8}_{-2.4}$} & $-2.1$ & - & - \\
& $\Sigma_c \bar{D}_s$-$\Xi_c' \bar{D}$ (D)
& $1$($\frac{1}{2}^-$) & $-1$ & \corr{$4446.1^{+1.9}_{-2.0(V)} - (0.1^{+0.4}_{-0.1(V)})\,i$} & $+0.3^{(*)}$ & - & - \\
& $\Xi_c \bar{D}^*$ & $1$($\frac{1}{2}^-$) & $-1$ & $(4478.1_{-2.1}^{+0.0} - 0.2_{-0.2(R)}^{+11.7}\,i)^V$ & $+0.5$ & - & - \\
\hline \hline
\multirow{4}{*}{(6) $\Xi_c \bar{D}^*$-$\Sigma_c^* \bar{D}_s^*$-$\Xi_c^* \bar{D}$}
& $\Xi_c \bar{D}^*$
& $1$($\frac{3}{2}^-$) & $-1$ & \corr{$4470.8^{+6.6}_{-4.4}$} & $-6.8$ & - & - \\
& $\Sigma_c^* \bar{D}_s$-$\Xi_c^* \bar{D}$ (O)
& $1$($\frac{3}{2}^-$) & $-1$ & \corr{$(4487.1^{+5.8}_{-2.0(R)} - (0.3^{+11.9}_{-0.3(R)})\,i)^V$} & $+5.5^{(*)}$ & - & - \\
& $\Sigma_c^* \bar{D}_s$-$\Xi_c^* \bar{D}$ (D) & $1$($\frac{3}{2}^-$) & $-1$ & \corr{$(4513.4^{+0.9}_{-2.1(R)} - (0.5_{-0.5(R)}^{+13.5})\,i)^V$} & $+0.4$ & - & - \\
\hline \hline
\multirow{4}{*}{(7) $\Xi_c' \bar{D}_s$-$\Omega_c \bar{D}$-$\Xi_c\bar{D}_s^*$}
& $\Xi_c' \bar{D}_s$-$\Omega_c \bar{D}$ (O)
& $\frac{1}{2}$($\frac{1}{2}^-$) & $-2$ & \corr{$4540.2^{+6.2}_{-5.0}$} & $-0.7$ & - & - \\
& $\Xi_c' \bar{D}_s$-$\Omega_c \bar{D}$ (D)
& $\frac{1}{2}$($\frac{1}{2}^-$) & $-2$ & \corr{$4558.6^{+7.6}_{-2.5} - (3.0^{+7.3}_{-2.2})\,i$} & $-1.6$ & - & - \\
& $\Xi_c\bar{D}_s^*$
& $\frac{1}{2}$($\frac{1}{2}^-$) & $-2$ & $4580.9^{+5.0}_{-3.6(V)} - (1.1^{+0.1}_{-1.1(V)})\,i$ & $+1.7$ & - & - \\
\hline \hline
\multirow{4}{*}{(8) $\Xi_c\bar{D}_s^*$-$\Xi_c^* \bar{D}_s$-$\Omega_c^* \bar{D}$}
& $\Xi_c\bar{D}_s^*$
& $\frac{1}{2}$($\frac{3}{2}^-$) & $-2$ & $4576.2^{+4.7}_{-1.5}$ & $-3.0$ & - & - \\
& $\Xi_c^* \bar{D}_s$-$\Omega_c^* \bar{D}$ (O)
& $\frac{1}{2}$($\frac{3}{2}^-$) & $-2$ & \corr{$4611.4^{+11.3}_{-3.0} - (1.5^{+1.5}_{-1.4})\,i$} & $+1.4$ & - & - \\
& $\Xi_c^* \bar{D}_s$-$\Omega_c^* \bar{D}$ (D)
& $\frac{1}{2}$($\frac{3}{2}^-$) & $-2$ & \corr{$4631.6^{+6.8}_{-3.6}- (1.9^{+0.1}_{-1.6})\,i$} & $+1.0$ & - & - \\
\hline \hline
\end{tabular}
\caption{Impact of the coupled channel dynamics of
  Eqs.~(\ref{eq:V-CC-3}-\ref{eq:V-CC-10})
  on the molecular pentaquark spectrum.
  ``Channels'' indicate the couple channels we are including
  in the calculation,  ``System'' the charmed antimeson-baryon
  we are dealing with, $I(J^P)$ the isospin, spin and parity of
  the system, $S$ the strangeness, $M^{\rm CC}_{\rm mol}$ the mass of
  the molecule, $\Delta M^{\rm CC}_{\rm mol}$  is how much the mass of the two-body
  state changes with respect to the single channel or two channel calculation
  that we previously considered in Tables
  \ref{tab:predictions-triplet-antitriplet},
  \ref{tab:predictions-octet}
  and \ref{tab:predictions-decuplet} (it is calculated
  for the central values only),
  ``Candidate'' refers to an experimentally observed pentaquark that could
  correspond with our theoretical prediction and
  $M_{\rm candidate}$ to its mass.
  In the column ``System'' the letters in parentheses --- $O$ and $D$ ---
  are used to specify whether the state is an octet or decuplet.
  A complex value of the mass indicates a resonance located in the
  $(II,I,I)$ or $(II,II,I)$ Riemann sheet (depending on whether
  it is the middle or the higher mass state, respectively).
  Meanwhile, if we use the superscript $V$ the pole will be in the
  $(I,II,II)$ or $(I,I,II)$ sheet.
  Changes of sheet owing to uncertainties in the location of the state
  are signaled with ``$(V)$'' or ``$(R)$'' attached to
  the lower errors.
  If the pole has changed sheet as a consequence of the coupled channel
  dynamics, we indicate it with a $(*)$ superscript
  in $\Delta M^{\rm CC}_{\rm mol}$.
  Masses are expressed in units of ${\rm MeV}$.
}
\label{tab:predictions-coupled-channels}
\end{table*}

The previous predictions can be further improved by including
the remaining potentially relevant coupled channel dynamics.
The impact of these coupled channel effects is more limited though: they
always involve the $\bar{H}_c T_c$ configurations, which are on average
lighter than the $\bar{H}_c S_c$ ones.
Thus, only a fraction of the $\bar{H}_c S_c$ predictions will be affected.

For what coupled channel dynamics to include, we will follow the detailed
analysis of Ref.~\cite{Yan:2021nio}, which quantified the size of
coupled channel contributions. 
Basically there will be eight cases to consider:
\begin{itemize}
\item[(1)] The $J=\tfrac{1}{2}$
  $\bar{D}^* \Lambda_c$-$\bar{D} \Sigma_c$ potential
  \begin{eqnarray}
    V(\bar{D}^* \Lambda_c-\bar{D} \Sigma_c) =
  \begin{pmatrix}
    \tilde{V}^{O} & -W^{O} \\
    -W^{O} & V^{O}
  \end{pmatrix} \, , \nonumber \\ \label{eq:V-CC-3}
  \end{eqnarray}
  where $\tilde{V}^O$ and $V^O$ refer to the octet potentials
  for the $\bar{H}_c T_c$ and $\bar{H}_c S_c$ type of molecular
  pentaquarks, while $W^{O}$ denotes the $\bar{H}_c T_c \to \bar{H}_c S_c$
  transition potential (which is always octet).
\item[(2)] The $J=\tfrac{1}{2}$ $\bar{D}_s \Lambda_c$-$\bar{D} \Xi_c$ potential
    \begin{eqnarray}
    && V(\bar{D}_s \Lambda_c-\bar{D} \Xi_c) = \nonumber \\ && \qquad
  \begin{pmatrix}
    \frac{1}{3} \tilde{V}^{S} + \frac{2}{3} \tilde{V}^{O}
    & \frac{\sqrt{2}}{3}\,(\tilde{V}^S - \tilde{V}^O) \\
    \frac{\sqrt{2}}{3}\,(\tilde{V}^S - \tilde{V}^O) &
    \frac{2}{3} \tilde{V}^{S} + \frac{1}{3} \tilde{V}^{O}
  \end{pmatrix} \, , \nonumber \\ \label{eq:V-CC-4}
    \end{eqnarray}
    where $\tilde{V}^S$ indicates the singlet potential
    for the $\bar{H}_c T_c$ molecules.
  \item[(3)] The $J=\tfrac{1}{2}$
    $\bar{D}_s^* \Lambda_c$-$\bar{D} \Xi_c'$-$\bar{D}^* \Xi_c$ potential
        \begin{eqnarray}
          && V(\bar{D}_s^* \Lambda_c-\bar{D}\Xi_c-\bar{D}^* \Xi_c) = \nonumber \\
          && \qquad
  \begin{pmatrix}
    \frac{1}{3} \tilde{V}^{S} + \frac{2}{3} \tilde{V}^{O}
    & -\sqrt{\frac{2}{3}} W^O &
    \frac{\sqrt{2}}{3}\,(\tilde{V}^S - \tilde{V}^O) \\
    -\sqrt{\frac{2}{3}} W^O & V^O & \frac{1}{\sqrt{3}} W^O \\
    \frac{\sqrt{2}}{3}\,(\tilde{V}^S - \tilde{V}^O) & \frac{1}{\sqrt{3}} W^O &
    \frac{2}{3} \tilde{V}^{S} + \frac{1}{3} \tilde{V}^{O}
  \end{pmatrix} \, , \nonumber \\ \label{eq:V-CC-5}
        \end{eqnarray}
  \item[(4)] The $J=\tfrac{3}{2}$
    $\bar{D}_s^* \Lambda_c$-$\bar{D}^* \Xi_c$-$\bar{D} \Xi_c^*$ potential
        \begin{eqnarray}
    && V(\bar{D}_s^* \Lambda_c-\bar{D}^* \Xi_c-\bar{D}\Xi_c^*) = \nonumber \\ && \qquad
  \begin{pmatrix}
    \frac{1}{3} \tilde{V}^{S} + \frac{2}{3} \tilde{V}^{O}
    & \frac{\sqrt{2}}{3}\,(\tilde{V}^S - \tilde{V}^O) &
    -\sqrt{\frac{2}{3}} W^O \\
    \frac{\sqrt{2}}{3}\,(\tilde{V}^S - \tilde{V}^O) &
    \frac{2}{3} \tilde{V}^{S} + \frac{1}{3} \tilde{V}^{O} &
    \frac{1}{\sqrt{3}} W^O  \\
    -\sqrt{\frac{2}{3}} W^O & \frac{1}{\sqrt{3}} W^O  & V^O 
  \end{pmatrix} \, , \nonumber \\ \label{eq:V-CC-6}
        \end{eqnarray}
   \item[(5)] The $J=\tfrac{1}{2}$
     $\bar{D}_s \Sigma_c$-$\bar{D} \Xi_c'$-$\bar{D}^* \Xi_c$ potential
      \begin{eqnarray}
        && V(\bar{D}_s \Sigma_c-\bar{D} \Xi_c'-\bar{D}^* \Xi_c) =
        \nonumber \\
          && \qquad
       \begin{pmatrix}
         \frac{2}{3} V^{O} + \frac{1}{3} V^{D} & - \frac{\sqrt{2}}{3}(V^O - V^D)
         & \sqrt{\frac{2}{3}} W^O \\
         - \frac{\sqrt{2}}{3}(V^O - V^D) & \frac{1}{3} V^{O} + \frac{2}{3} V^{D}
         & -\frac{1}{\sqrt{3}} W^O  \\
         \sqrt{\frac{2}{3}} W^O & -\frac{1}{\sqrt{3}} W^O  & \tilde{V}^O
       \end{pmatrix} \, , \nonumber \\ \label{eq:V-CC-7}
     \end{eqnarray}
   \item[(6)] The $J=\tfrac{3}{2}$
     $\bar{D}^* \Xi_c$-$\bar{D}_s^* \Sigma_c$-$\bar{D} \Xi_c^*$ potential
      \begin{eqnarray}
        && V(\bar{D}^* \Xi_c-\bar{D}_s^* \Sigma_c-\bar{D} \Xi_c^*) =
        \nonumber \\
          && \qquad
       \begin{pmatrix}
       \tilde{V}^O &  \sqrt{\frac{2}{3}} W^O & -\frac{1}{\sqrt{3}} W^O  \\ 
       \sqrt{\frac{2}{3}} W^O & \frac{2}{3} V^{O} + \frac{1}{3} V^{D}
       & - \frac{\sqrt{2}}{3}(V^O - V^D) \\
       -\frac{1}{\sqrt{3}} W^O  & - \frac{\sqrt{2}}{3}(V^O - V^D) &
       \frac{1}{3} V^{O} + \frac{2}{3} V^{D} \\
     \end{pmatrix} \, , \nonumber \\ \label{eq:V-CC-8}
     \end{eqnarray}
   \item[(7)] The $J=\tfrac{1}{2}$
     $\bar{D}_s \Xi_c'$-$\bar{D} \Omega_c$-$\bar{D}_s^* \Xi_c$ potential
     \begin{eqnarray}
       && V(\bar{D}_s \Xi_c^{'}-\bar{D} \Omega_c-\bar{D}_s^* \Xi_c) =
       \nonumber \\ && \qquad
     \begin{pmatrix}
       \frac{1}{3} V^{O} + \frac{2}{3} V^{D} & - \frac{\sqrt{2}}{3}(V^O - V^D)
       & \frac{1}{\sqrt{3}} W^O  \\ 
       - \frac{\sqrt{2}}{3}(V^O - V^D) & \frac{2}{3} V^{O} + \frac{1}{3} V^{D}
       & -\sqrt{\frac{2}{3}} W^O \\
       \frac{1}{\sqrt{3}} W^O  & -\sqrt{\frac{2}{3}} W^O & \tilde{V}^O
     \end{pmatrix} \, , \nonumber \\ \label{eq:V-CC-9}
     \end{eqnarray}
   \item[(8)] The $J=\tfrac{1}{2}$
     $\bar{D}_s^* \Xi_c$-$\bar{D}_s \Xi_c^*$-$\bar{D} \Omega_c^*$ potential
     \begin{eqnarray}
       && V(\bar{D}_s^* \Xi_c-\bar{D}_s \Xi_c^{*}-\bar{D} \Omega_c^*) =
       \nonumber \\ && \qquad
       \begin{pmatrix}
         \tilde{V}^O & \frac{1}{\sqrt{3}} W^O  & -\sqrt{\frac{2}{3}} W^O \\  
       \frac{1}{\sqrt{3}} W^O & \frac{1}{3} V^{O} + \frac{2}{3} V^{D}
       & - \frac{\sqrt{2}}{3}(V^O - V^D) \\ 
       -\sqrt{\frac{2}{3}} W^O & - \frac{\sqrt{2}}{3}(V^O - V^D)
       & \frac{2}{3} V^{O} + \frac{1}{3} V^{D} 
     \end{pmatrix} \, . \nonumber \\ \label{eq:V-CC-10}
     \end{eqnarray}
\end{itemize}

For the $\bar{H}_c T_c$ molecules, the singlet and octet components of
the potential are
\begin{eqnarray}
  \tilde{V}^S  = \tilde{C}^S &\propto& 
  -4\,\frac{g_{V1} g_{V2}}{m_V^2}
  - {(\frac{m_V}{m_S})}^{\alpha}\frac{g_{S1} g_{S2}}{m_S^2} \, , \\
  \tilde{V}^O = \tilde{C}^O &\propto& 
  +2\,\frac{g_{V1} g_{V2}}{m_V^2}
  - {(\frac{m_V}{m_S})}^{\alpha}\frac{g_{S1} g_{S2}}{m_S^2} \, ,
\end{eqnarray}
where in this case there is no light-spin dependence.
The $\bar{H}_c T_c$-$\bar{H}_c S_c$ transition component is
\begin{eqnarray}
  W^O = E^O &\propto& 
  -2\,\sqrt{3}\,\frac{g_{V1} g_{V2}}{m_V^2}
  \kappa_{V1} \kappa_{V2} \, \frac{m_V^2}{6 M^2}\,\hat{C}_{L12} \, ,
\end{eqnarray}
which is purely magnetic-like ($M1$), with
$\kappa_{V2} = \frac{3}{2}\,({\mu_u}/{\mu_N})$
for the $\Xi_c \to \Xi_c^{('/*)}$ transitions~\cite{Peng:2020hql},
and where the value of the spin-spin operator always happens
to be one, $\hat{C}_{L12} = 1$.

For the calculation of the spectrum with the extended coupled channels,
we will use the $P_{\psi}^N(4312)$ as the reference state,
but still in the single channel approximation.
The reason for not updating the input to its coupled channel version is
the following:
if we recalculate the $P_{\psi}^N(4312)$ with the coupled channel dynamics
of Eq.~(\ref{eq:V-CC-3}) and the single channel reference coupling
$C^{\rm sat}_{\rm ref} = -0.80^{+0.14}_{-0.01}\,{\rm fm}^2$,
we now predict the mass of the $P_{\psi}^N(4312)$ to be
\begin{eqnarray}
  M(\bar{D} \Sigma_c) = 4315.8^{+4.8}_{-2.3} - (6.6^{+8.2}_{-4.2})
  \, i \, {\rm MeV} \, ,
\end{eqnarray}
which its compatible with the experimental mass and width of
the $P_{\psi}^N(4312)$ within errors
($M_{\rm exp} - (\Gamma_{\rm exp}/2)\,i = [4311.9^{+6.8}_{-0.9} - (4.9^{+2.3}_{-2.6})\,i]\,{\rm MeV}$).
As a consequence it is not necessary (within the accuracy of the theory
or the experiment) to recalibrate the input.

With the previous choice, the predictions of the states affected
by the extended coupled channel dynamics of
Eqs.~(\ref{eq:V-CC-3}-\ref{eq:V-CC-10})
are presented in Table \ref{tab:predictions-coupled-channels}.
Within this Table we include the calculation of the mass shift generated
by the coupled channel dynamics,
$\Delta M^{\rm CC}_{\rm mol}$ in Table \ref{tab:predictions-coupled-channels}.
It is interesting to notice that this mass shift is in general smaller
in magnitude than the uncertainties we previously calculated
in Tables \ref{tab:predictions-triplet-antitriplet},
\ref{tab:predictions-octet} and \ref{tab:predictions-decuplet}.
There are exceptions though to this rule: the most notable one is
the $I=1$, $J=\tfrac{3}{2}$ $\Xi_c \bar{D}^*$ state
--- a $P_{\psi s}^{\Sigma}$ pentaquark ---
which becomes about $7\,{\rm MeV}$ more bound than
in the single channel calculation.
Other example is the $I=0$, $J=\tfrac{1}{2}$ $\Lambda_c \bar{D}^*$ state
--- a $P_{\psi}^N$ pentaquark --- which before was a virtual state
basically at threshold and now is a bound state a few ${\rm MeV}$
below threshold.
Regarding the $P_{\psi s}^{\Lambda}(4459)$ pentaquark, the inclusion of the
coupled channel dynamics generates a certain degree of hyperfine
splitting between the $J = \tfrac{1}{2}$ and $\tfrac{3}{2}$
$\Xi_c \bar{D}^*$ configurations ($| \Delta M | = 4.2\,{\rm MeV}$),
though of a smaller magnitude than the double peak solution
in~\cite{Aaij:2020gdg} ($| \Delta M_{\rm exp} | = 12.9 \pm 4.6\,{\rm MeV}$).

\subsection{Isospin breaking effects}

\begin{table*}[!ttt]
\begin{tabular}{|clllllc|}
\hline\hline
System  & $I$($J^{P}$) & $M_{\rm mol}$ & $I$($J^{P}$) & $M_{\rm mol}$ &
Candidate & $M_{\rm candidate}$ \\
\hline
$\Xi_c^0 \bar{D}^0$-$\Xi_c^+ D^-$ & $0$ ($\tfrac{1}{2}^{-}$) &  $4327.3^{+6.7}_{-0.9}$ &
$1$ ($\tfrac{1}{2}^{-}$) &  $4337.3^{+4.7}_{-2.3} - (0.9^{+0.0}_{-0.9(B/V)})\,i$ &
$P_{\psi s}^{\Lambda}(4338)$ & $4338.2 \pm 0.7$ \\
\hline \hline
\multirow{2}{*}{$\Sigma_c^{++} D^-$-$\Sigma_c^+ \bar{D}^0$} &
\multirow{2}{*}{$\tfrac{1}{2}$ ($\tfrac{1}{2}^{-}$)} &
\multirow{2}{*}{$4312.2^{+5.3}_{-0.9}$} &
\multirow{2}{*}{$\tfrac{3}{2}$ ($\tfrac{1}{2}^{-}$)} &
\multirow{2}{*}{$4325.6^{+7.1}_{-4.1} - (3.9^{+4.8}_{-2.4})\, i$} &
$P_{\psi}^{N}(4312)$ & $4311.9^{+6.8}_{-0.9}$ \\
& & & & & $P_{\psi}^{N}(4337)$ & $4337^{+7}_{-5}$ \\
$\Sigma_c^{*++} D^-$-$\Sigma_c^{*+} \bar{D}^0$ &
$\tfrac{1}{2}$ ($\tfrac{3}{2}^{-}$) &
$4376.3^{+5.9}_{-0.9}$ &
$\tfrac{3}{2}$ ($\tfrac{3}{2}^{-}$) &
$4390^{+7.4}_{-4.3} - (4.0^{+4.9}_{-2.2})\, i$ &
- & - \\
$\Sigma_c^{++} D^{*-}$-$\Sigma_c^{+} \bar{D}^{*0}$ &
$\tfrac{1}{2}$ ($\tfrac{1}{2}^{-}$) &
$4461.1^{+3.0}_{-2.5} - (0.2^{+0.2}_{-0.2(B)})\, i$ &
$\tfrac{3}{2}$ ($\tfrac{1}{2}^{-}$) &
$4453.9^{+4.9}_{-1.6}$ &
$P_{\psi}^N(4457)$ & $4457.3^{+4.1}_{-1.8}$ \\
$\Sigma_c^{*++} D^{*-}$-$\Sigma_c^{*+} \bar{D}^{*0}$ &
$\tfrac{1}{2}$ ($\tfrac{1}{2}^{-}$) &
$(4526.3 \pm 2.7) - (0.5^{+0.2}_{-0.5(B)})\, i$ &
$\tfrac{3}{2}$ ($\tfrac{1}{2}^{-}$) &
$4513.3^{+8.1}_{-1.8}$ &
- & - \\
$\Sigma_c^{*++} D^{*-}$-$\Sigma_c^{*+} \bar{D}^{*0}$ &
$\tfrac{1}{2}$ ($\tfrac{3}{2}^{-}$) &
$4520.1^{+4.1}_{-1.9}$ &
$\tfrac{3}{2}$ ($\tfrac{3}{2}^{-}$) &
$4525.9^{+2.6}_{-2.5} - (0.7^{+0.8}_{-0.7(B)})\, i$ &
- & - \\
\hline\hline
\end{tabular}
\caption{Selection of molecular pentaquark states for which the isospin
  breaking effects are worth considering.
  ``System'' indicates which charmed antimeson-baryon we are dealing with,
  $I(J^P)$ the isospin, spin and parity of the molecule, $R_{\rm mol}$ the
  molecular ratio (the relative attractiveness of the molecule relative
  to the $P_{\psi}^N(4312)$) as defined in Eq.~(\ref{eq:Rmol}),
  $M_{\rm mol}$ the mass of the molecule,   ``Candidate'' refers to
  an experimentally observed pentaquark that could
  correspond with our theoretical prediction and
  $M_{\rm candidate}$ to its mass.
  Masses are expressed in units of ${\rm MeV}$.
  In the $M_{\rm mol}$ column the superscript ``$V$'' indicates a virtual state.
  The errors in $M_{\rm mol}$ are calculated
  as in Table \ref{tab:predictions-triplet-antitriplet}.
  When the uncertainties allow a state to change from virtual to bound or
  vice versa, we indicate it by adding $(V)$ or $(B)$ after
  the upper error.
}
\label{tab:predictions-isospin-breaking}
\end{table*}

Within the previous set of predictions it is not rare to find a few
states close to a threshold, opening the possibility of isospin
breaking effects when the relevant threshold can be decomposed
into two nearby particle states with the same third
component of their total isospin.
Examples are the $P_{\psi s}^{\Sigma}(4336)$ molecule predicted
in Table \ref{tab:predictions-triplet-antitriplet}
or the $P_{\psi}^N(4460)$ and $P_{\psi}^{\Delta}(4456)$
molecules of Tables \ref{tab:predictions-octet}
and \ref{tab:predictions-decuplet},
which are  experimentally relevant owing to their possible
relation with the $P_{\psi s}^{\Lambda}(4338)$ and $P_{\psi}^N(4457)$ states.

Here we consider the isospin breaking of the thresholds for the particular
cases of the $\Xi_c \bar{D}$ system and a few selected configurations
of the $\Sigma_c^{(*)} \bar{D}^{(*)}$ molecules.
The inclusion of these effects is straightforward:
\begin{itemize}
\item[(a)] First, we split the $\Xi_c \bar{D}$ and $\Sigma_c^{(*)} \bar{D}^{(*)}$
  systems into the $\Xi_c^0 \bar{D}^0$-$\Xi_c^+ D^-$ and
  $\Sigma_c^{(*)+} \bar{D}^{(*0)}$-$\Sigma_c^{(*)++} {D}^{(*-)}$
  channels, respectively.
\item[(b)] Second, we substitute the isospin operator by
  \begin{eqnarray}
    T_{12} &\to&
    \begin{pmatrix}
      -1 & +2 \\
      +2 & - 1
    \end{pmatrix} \quad \mbox{for $\Xi_c^0 \bar{D}^0$-$\Xi_c^+ D^-$,}
    \\
    T_{12} &\to&
    \begin{pmatrix}
      0 & +\sqrt{2} \\
      +\sqrt{2} & - 1
    \end{pmatrix} \quad \mbox{for $\Sigma_c^{(*)+} \bar{D}^{(*0)}$-$\Sigma_c^{(*)++} {D}^{(*-)}$,}
  \end{eqnarray}
  in Eq.~(\ref{eq:coupling-sat}), which now becomes a coupled channel
  equation, while every other operator remains diagonal
  in the particle basis.
\end{itemize}
Then we recalculate the spectrum
in Table \ref{tab:predictions-isospin-breaking},
where we classify states into approximate isospin representations by
adapting the criterion we already used for SU(3)-flavor
breaking to isospin.
That is, we inspect the vertex functions $\phi_A$ (with $A = 1,2$) of
the states that are below threshold in the particle basis and
compare them with their expected value
in the absence of isospin breaking.
For instance, in the $\Xi_c^0 \bar{D}^0$-$ \Xi_c^+ D^-$ system we obtain
two poles with masses and vertex functions:
\begin{eqnarray}
  M_L &=& 4327.4\,{\rm MeV} \, , \\
  {(\phi_1,\phi_2)}_L &=& (0.74,-0.68) \, , \\
  \nonumber \\
  M_H &=& (4337.3 - 0.9\,i)\,{\rm MeV}\, ,  \\
  {(\phi_1,\phi_2)}_H &=& (0.35-0.40\,i, 0.85) \, ,
\end{eqnarray}
for the lower and higher mass poles (subscripts $L$ and $H$), respectively.
In the isospin symmetric limit, the $\Xi_c \bar{D}$ system can be
in the $I=0, 1$ configurations, where for $I=0$ we expect the vertex
functions to be
$(\phi_1, \phi_2) = (1/\sqrt{2},-1/\sqrt{2}) \simeq (0.71, -0.71)$
in the particle basis.
From this we conclude that the lower mass pole is predominantly $I=0$
and classify the higher mass pole as $I=1$.

The spectrum of Table \ref{tab:predictions-isospin-breaking} is worth
commenting in more detail owing to its possible connection
with experimental observations:
\begin{itemize}
\item[(i)] In the $\Xi_c \bar{D}$ system, the mass of the $I=1$ state is
  really close to the one of the $P_{\psi s}^{\Lambda}(4338)$ pentaquark,
  which begs the question of whether this peak is actually a
  $P_{\psi s}^{\Sigma^0} $ or a combination of the two peaks
  we predict.
  This is not impossible if we take into account that
  isospin breaking effects allow both the $P_{\psi s}^{\Lambda}$ and
  $P_{\psi s}^{\Sigma^0}$ to decay into $J/\Psi \Lambda$ (instead of
  only $P_{\psi s}^{\Lambda}$ if isospin is not broken).
  
  However, the analysis of Ref.~\cite{Nakamura:2022jpd}, which includes
  both the $\Xi_c^0 \bar{D}^0$ and $\Xi_c^+ {D}^-$ channels, prefers
  a $P_{\psi s}^{\Lambda}$-type solution (i.e. $I=0$), which
  happens to be above both these thresholds.
  This corresponds to a $\Xi_c \bar{D}$ system somewhat less attractive
  than what we find in our model.

\item[(ii)] In the $\Sigma_c \bar{D}$ system the $I=3/2$ $P_{\psi}^{\Delta}$
  molecule, which was a virtual state in the isospin symmetric limit,
  is now a resonance above the $\Sigma_c^+ \bar{D}^0$ and
  $\Sigma_c^{++} {D}^-$ thresholds.
  Its mass is similar to that of the $P^N_{\psi}(4337)$~\cite{Aaij:2021august},
  a pentaquark whose nature remains obscure~\cite{Yan:2021nio,Nakamura:2021dix}.
  This similarity in the masses invites the re-interpretation of
  the $P^N_{\psi}(4337)$ as a possible $P_{\psi}^{\Delta}$ molecular
  pentaquark, instead of the less exotic
  $P_{\psi}^N$ interpretation.
  However, this explanation is contingent on two factors:
  first, that the spin and parity of the $P^N_{\psi}(4337)$ are
  $J^P = \tfrac{1}{2}^-$ (instead of the preferred
  $J^P = \tfrac{1}{2}^+$~\cite{Aaij:2021august}) and
  second, that there is destructive interference in
  the $B_s^0 \to J/\Psi p \bar{p}$ between the contributions
  from the $P^N_{\psi}$ and $P^{\bar{N}}_{\psi}$ -- i.e. the $I=1/2$
  $\Sigma_c \bar{D}$ and $\bar{\Sigma}_c D$ molecular pentaquarks --
  while the contributions from the $I=3/2$ $P^{\Delta}_{\psi}$ and
  $P^{\bar{\Delta}}_{\psi}$ states interfere constructively.
  As a reminder, it is interesting to notice that the experimental analysis
  from which the properties of the $P_{\psi}^N$ were determined assumes
  the existence of just one pentaquark in the vicinity of
  the $4337\,{\rm MeV}$ mass range~\cite{Aaij:2021august}.
  
\item[(iii)] For the $J=1/2$ $\Sigma_c \bar{D}^*$ system there was already
  more attraction in the $I=3/2$ configuration than in the $I=1/2$ one
  in the isospin symmetric limit.
  The inclusion of isospin breaking effects makes the $P_{\psi}^{\Delta}$ a bit
  more bound, while pushing the $P_{\psi}^N$ above
  the $\Sigma_c^+ \bar{D}^{0*}$ threshold.
  This is relevant for the interpretation of the $P_{\psi}^+(4457)$ pentaquark,
  which is usually considered to be $I=1/2$ and have the quantum
  numbers of a proton.
  But once isospin breaking is taken into account, its interpretation as
  a mostly $I=3/2$ $P_{\psi}^{\Delta}$ state is more natural
  from the point of view of spectroscopy~\footnote{Had we considered
    the effect of the nearby $\Lambda_{c1}(2595) \bar{D}$ threshold
    (a possibility which has been discussed
    in the literature~\cite{Burns:2015dwa,Geng:2017hxc,Burns:2019iih,Peng:2020gwk,Yalikun:2021bfm}),
    with a mass of $4459.5$ and thus below the prediction of
    the $J=1/2$, $I=1/2$ $\Sigma_c \bar{D}^*$ in Table
    \ref{tab:predictions-isospin-breaking} (i.e. $4461.1\,{\rm MeV}$),
    this effect would have further increased the mass of
    the $P_{\psi}^N$, making it a less likely interpretation of
    the $P_{\psi}^N(4457)$.
    However, the $\Sigma_c \bar{D}^*$-$\Lambda_{c1}(2595) \bar{D}$ dynamics
    involve P-wave interactions, for which our saturation model does not
    apply (as it is constrained to the S-wave case).
  }.
  Yet, this is not the only reason in favor of the $I=3/2$ hypothesis:
  this option also explains better the {\it production fractions} of the
  LHCb pentaquarks~\footnote{Here by production fractions we refer
    to the following ratio of branching ratios:
    \begin{eqnarray}
      \mathcal{F}(P_{\psi}) =
      \frac{\mathcal{B}(\Lambda_b^0 \to K^- P_{\psi}^+)\,
        \mathcal{B}(P_{\psi}^+ \to J/\Psi p)}{\mathcal{B}
        (\Lambda_b^0 \to K^- J/\Psi p)} \, ,
    \end{eqnarray}
    where $\mathcal{B}$ are the branching ratios of the particular decays
    considered. Experimentally, for the $P_{\psi}^+(4312/4440/4457)$
    pentaquarks we have ${\mathcal F} = 0.30^{+0.35}_{-0.11}$,
    $1.11^{+0.40}_{-0.34}$, $0.53^{+0.22}_{-0.21}$, respectively~\cite{Aaij:2019vzc}
    (notice that we have removed the $^N$ superscripts in the pentaquarks
    to indicate that we are not making any assumptions
    about their isospin).
    Alternatively, if we want to compare the relative production ratios,
    we might divide by $\mathcal{F}(P_{\psi}(4312))$, yielding
    $1$, $3.7^{+2.5}_{-2.3}$ and $1.8 \pm 1.2$.
  }.
  If the $P_{\psi}(4457)$ were to be a $J=1/2$, $I=1/2$ $\Sigma_c \bar{D}$
  molecule, its production fraction could be expected to be one order of
  magnitude larger than the experimental
  one~\cite{Sakai:2019qph,Peng:2022iez}.
  In contrast, with the spectrum obtained in Table \ref{tab:predictions-octet}
  for the $P_{\psi}^N(4440)$ and Table \ref{tab:predictions-isospin-breaking}
  for the $P_{\psi}^N(4312)$ and the prospective $P_{\psi}^{\Delta}(4457)$,
  the calculation of the relative production fractions (check
  Ref.~\cite{Peng:2022iez} for a detailed explanation) yields
  \begin{eqnarray}
    1 : 1.67 \, \mathcal{R}_2 : 0.86 \, \mathcal{R}_3 \,\, [11.5\,\mathcal{R}_3] \, ,
  \end{eqnarray}
  for $P_{\psi}^N(4312)$, $P_{\psi}^N(4440)$ and $P_{\psi}^{\Delta}(4457)$
  [$P_{\psi}^N(4457)$], respectively, where $\mathcal{R}_2$ and
  $\mathcal{R}_3$ are the ratios of the $\mathcal{B}(\Lambda_b^0 \to K^- P_c^+)$
  branching ratios for the $P_c^+ = P_{\psi}^N(4440)$ and
  $P_{\psi}^{\Delta}(4457)$ [$P_{\psi}^N(4457)$]
  relative to $P_{\psi}^N(4312)$.
  In the calculation above we are not using the physical pentaquark masses,
  but instead their predictions within our RG-saturation model:
  the $P_{\psi}^N(4312)$ and $P_{\psi}^{\Delta}(4457)$ are the ones calculated
  in Table \ref{tab:predictions-isospin-breaking}, while the
  $P_{\psi}^N(4440)$ corresponds to Table \ref{tab:predictions-octet}.
  The number within the square brackets represents the ratio 
  for the $P_{\psi}^N(4457)$ as calculated from the mass of the molecular
  prediction in Table \ref{tab:predictions-octet}.
  For comparison, the experimental values are~\cite{Aaij:2019vzc}
  \begin{eqnarray}
    1 : 3.7^{+2.5}_{-2.3} : (1.8 \pm 1.2) \, ,
  \end{eqnarray}
  which are compatible with our determination provided $\mathcal{R}_2$
  and $\mathcal{R}_3$ are of natural size.
  In contrast, for the $P_{\psi}^N(4457)$ of Table \ref{tab:predictions-octet}
  the ratio $\mathcal{R}_3$ needs to be rather small.
  
\item[(iv)] For the $\Sigma_c^* \bar{D}^*$ system, we find two
  states above the $\Sigma_c^+ \bar{D}^{0*}$ threshold:
  a $J=1/2$ $P_{\psi}^N$ and $J=3/2$ $P_{\psi}^{\Delta}$.
  The $I=3/2$ molecule might in turn be related with a possible
  $I=3/2$ structure with a mass in the $(4545-4553)\,{\rm MeV}$ range
  recently observed in the $\Lambda_c^+ D^- \pi^-$
  decay channel by the LHCb~\cite{LHCb:2024pnt} (though with a poor
  local significance, ranging from $(1.9-3.6)\,{\sigma}$ depending
  on the assumed width).
\end{itemize}
We remind that the present analysis can be easily extended to other
pentaquarks close to a threshold, which could come in handy
if future experimental observations find
this type of states.
For the moment, though, we have limited the explicit inclusion of isospin
breaking to the pentaquarks were it might be empirically relevant.

\subsection{Using the $P_{\psi s}^{\Lambda}(4338/4459)$ as the reference state}

\begin{table}[!ttt]
\begin{tabular}{|clllcl|}
\hline\hline
System  & $I$($J^{P}$) & $R_{\rm mol}'$ & $M_{\rm mol}'$ &
Candidate & $M_{\rm candidate}$\\
\hline
  \hline
  $\Lambda_c \bar{D}_s$ & $0$ ($\tfrac{1}{2}^{-}$) & $0.86$ & 
  $(4253.1^{+1.0}_{-2.3})^V$ & - & - \\
  $\Lambda_c \bar{D}_s^*$ & $0$ ($\tfrac{1}{2}^{-}$,$\tfrac{3}{2}^{-}$) & $0.89$ &
  $(4397.7^{+0.7}_{-1.8})^V$ & - & - \\
\hline
$\Xi_c \bar{D}$ & $0$ ($\tfrac{1}{2}^{-}$) & $1.00$ & 
$4336.3$ (Input) & $P_{\psi s}^{\Lambda}(4338)$ & $4338.2 \pm 0.7$ \\
$\Xi_c \bar{D}^*$ & $0$ ($\tfrac{1}{2}^{-}$,$\tfrac{3}{2}^{-}$) & $1.04$ & 
$4477.5^{+0.1}_{-0.2}$ & $P_{\psi s}^{\Lambda}(4459)$ & $4458.9^{+5.5}_{-3.1}$ \\
\hline
  $\Sigma_c \bar{D}$ & $\tfrac{1}{2}$ ($\tfrac{1}{2}^{-}$) & $1.00$ 
& $(4320.7 \pm 0.0)^V$ & $P_{\psi}^N(4312)$ & $4311.9^{+6.8}_{-0.9}$ \\
  $\Sigma_c^* \bar{D}$ & $\tfrac{1}{2}$ ($\tfrac{3}{2}^{-}$) & $1.01$ 
  & $4385.3 \pm 0.0$ & - & - \\
$\Sigma_c \bar{D}^*$ & $\tfrac{1}{2}$ ($\tfrac{1}{2}^{-}$) & $0.86$
& $(4460.3^{+1.2}_{-3.2})^V$ & $P_{\psi}^N(4457)$ & $4457.3^{+4.1}_{-1.8}$ \\
$\Sigma_c \bar{D}^*$ & $\tfrac{1}{2}$ ($\tfrac{3}{2}^{-}$) & $1.13$
& $4460.9^{+0.5}_{-1.5}$ & $P_{\psi}^N(4440)$  & $4440.3^{+4.3}_{-1.8}$ \\
  $\Sigma_c^* \bar{D}^*$ & $\tfrac{1}{2}$ ($\tfrac{1}{2}^{-}$) & $0.82$  
 & $(4524.0^{+2.0}_{-5.3})^V$ & -  & - \\
$\Sigma_c^* \bar{D}^*$ & $\tfrac{1}{2}$ ($\tfrac{3}{2}^{-}$) & $0.96$
& $(4526.5^{+0.1}_{-0.5})^V$ & -  & - \\
$\Sigma_c^* \bar{D}^*$ & $\tfrac{1}{2}$ ($\tfrac{5}{2}^{-}$) & $1.19$
& $4524.4^{+1.0}_{-2.8}$ & -  & - \\
\hline\hline 
\end{tabular}
\caption{Selection of molecular pentaquark states when
  the $P_{\psi s}^{\Lambda}(4338)$ is chosen as the reference state
  (with the caveat that we describe it as an isoscalar
  $\bar{D} \Xi_c$ bound state at threshold).
  We refer to Table \ref{tab:predictions-triplet-antitriplet}
  for the conventions we follow, where the only change is
  the addition of a prime to $R_{\rm mol}'$ and $M_{\rm mol}'$
  with the intention of indicating that the molecular ratios and
  masses are now derived from the $P_{\psi s}^{\Lambda}(4338)$.
  The uncertainties shown here only take into account the mass of
  the scalar meson, the cutoff choice and the parameter $\alpha$.
  All binding energies and masses are in units of ${\rm MeV}$.
}
\label{tab:predictions-Pcs4338-input}
\end{table}

\begin{table}[!ttt]
\begin{tabular}{|clllcl|}
\hline\hline
System  & $I$($J^{P}$) & $R_{\rm mol}''$ & $M_{\rm mol}''$ &
Candidate & $M_{\rm candidate}$\\
\hline
  \hline
  $\Lambda_c \bar{D}_s$ & $0$ ($\tfrac{1}{2}^{-}$) & $0.82$ & 
  $4248.5^{+3.4}_{-3.1}$ & - & - \\
  $\Lambda_c \bar{D}_s^*$ & $0$ ($\tfrac{1}{2}^{-}$,$\tfrac{3}{2}^{-}$) & $0.85$ &
  $4390.8^{+3.8}_{-3.4}$ & - & - \\
\hline
$\Xi_c \bar{D}$ & $0$ ($\tfrac{1}{2}^{-}$) & $0.96$ & 
$4320.2^{+5.1}_{-2.9}$ & $P_{\psi s}^{\Lambda}(4338)$ & $4338.2 \pm 0.7$ \\
$\Xi_c \bar{D}^*$ & $0$ ($\tfrac{1}{2}^{-}$,$\tfrac{3}{2}^{-}$) & $1.00$ & 
Input & $P_{\psi s}^{\Lambda}(4459)$ & $4458.9^{+5.5}_{-3.1}$ \\
\hline
  $\Sigma_c \bar{D}$ & $\tfrac{1}{2}$ ($\tfrac{1}{2}^{-}$) & $0.96$ 
& $4304.8^{+5.1}_{-2.9}$ & $P_{\psi}^N(4312)$ & $4311.9^{+6.8}_{-0.9}$ \\
  $\Sigma_c^* \bar{D}$ & $\tfrac{1}{2}$ ($\tfrac{3}{2}^{-}$) & $0.97$ 
  & $4368.7^{+5.2}_{-2.9}$ & - & - \\
$\Sigma_c \bar{D}^*$ & $\tfrac{1}{2}$ ($\tfrac{1}{2}^{-}$) & $0.82$
& $4455.9 \pm 4.0$ & $P_{\psi}^N(4457)$ & $4457.3^{+4.1}_{-1.8}$ \\
$\Sigma_c \bar{D}^*$ & $\tfrac{1}{2}$ ($\tfrac{3}{2}^{-}$) & $1.09$
& $4435.6^{+6.8}_{-4.2}$ & $P_{\psi}^N(4440)$  & $4440.3^{+4.3}_{-1.8}$ \\
  $\Sigma_c^* \bar{D}^*$ & $\tfrac{1}{2}$ ($\tfrac{1}{2}^{-}$) & $0.79$  
 & $4522.4^{+3.7}_{-4.5}$ & -  & - \\
$\Sigma_c^* \bar{D}^*$ & $\tfrac{1}{2}$ ($\tfrac{3}{2}^{-}$) & $0.92$
& $4514.4^{+4.3}_{-2.5}$ & -  & - \\
$\Sigma_c^* \bar{D}^*$ & $\tfrac{1}{2}$ ($\tfrac{5}{2}^{-}$) & $1.14$
& $4495.1^{+7.5}_{-6.8}$ & -  & - \\
\hline\hline 
\end{tabular}
\caption{Same as Table \ref{tab:predictions-Pcs4338-input} but
  using the $P_{\psi s}^{\Lambda}(4459)$ as a reference state,
  where the only change is the addition of a second prime
  to $R_{\rm mol}''$ and $M_{\rm mol}''$ to distinguish them
  from the molecular ratios and masses obtained
  from the other input choices.
  The uncertainties here include the error in the mass of the
  reference state, in addition to the other three error
  sources ($m_{\sigma}$, $\Lambda$, $\alpha$).
  All binding energies and masses are in units of ${\rm MeV}$.
}
\label{tab:predictions-Pcs4459-input}
\end{table}

Finally we consider the question of how predictions change
if a different reference state is chosen.
Ideally, the reference state should have a straightforward molecular
interpretation and, if possible, be also well-established
experimentally.
This leaves us with five possible candidates: the $P_{\psi}^N(4312/4440/4457)$
and the $P_{\psi s}^{\Lambda}(4338/4459)$ (though, of the strange pentaquarks,
only the $P_{\psi s}^{\Lambda}(4338)$ surpasses the $5\,\sigma$ discovery
threshold).
However, the three non-strange pentaquarks are known to be consistently
described with similar parameters~\cite{Liu:2019tjn,Guo:2019fdo}.
In terms of the RG-improved saturation model this manifests as the fact that
the predictions for the masses of the $P_{\psi}^N(4440)$ and
$P_{\psi}^N(4457)$ obtained from the $P_{\psi}^N(4312)$ are
compatible with the experimental ones, as can be checked
in Table \ref{tab:predictions-octet}.
In this case changing the reference state from the $P_{\psi}^N(4312)$ to any of
the other two will not significantly alter the predictions.

The situation is different though with the $P_{\psi s}^{\Lambda}(4338)$
and $P_{\psi s}^{\Lambda}(4459)$ pentaquarks, which overbinds and
underbinds respectively in our model if the
$P_{\psi}^N(4312)$ is used as input (see Table
\ref{tab:predictions-triplet-antitriplet}).
Thus, it is worth exploring how predictions change if we use
the $P_{\psi s}^{\Lambda}(4338/4459)$ pentaquarks
as the reference state.

The only difficulty is that if we use the $P_{\psi s}^{\Lambda}(4338)$ as input
its mass is above the $\Xi_c \bar{D}$ threshold, which is not reproducible
in a single channel calculation.
Actually, even after the inclusion of coupled channel effects, it is still
not possible to obtain the mass of the $P_{\psi s}^{\Lambda}(4338)$:
if we consider isospin breaking effects,
the $\Xi_c^0 \bar{D}^0$-$\Xi_c^+ {D}^-$ coupled channel dynamics is able to
generate an isovector $P_{\psi s}^{\Sigma}$ molecule above threshold,
but not an isoscalar $P_{\psi s}^{\Lambda}$ one.
Conversely, if we consider the $\Lambda_c \bar{D}_s$-$\Xi_c \bar{D}$ coupled
channel dynamics, the transition potential is able to generate a resonance
barely above threshold, but not above enough as to reproduce
the Breit-Wigner mass of the $P_{\psi s}^{\Lambda}(4338)$
(for that we will need a stronger transition potential than
the one generated by $K^*$-exchange in our model).

In view of the previous limitations, we will model
the $P_{\psi s}^{\Lambda}(4338)$ as a $I=1/2$ $\Xi_c \bar{D}$ bound state
at threshold, i.e. $M_{\rm th} = 4336.3\,{\rm MeV}$ in the isospin
symmetric limit (for which $C^{\rm sat}_{\rm mol} = -0.58\,{\rm fm}^2$
for $\Lambda = 1.0\,{\rm GeV}$).
To keep matters simple, we will concentrate on the most experimentally
relevant sectors: the $\Lambda_c \bar{D}_s^{(*)}$ (for which there are
predictions of a partner of
the $P_{\psi s}^{\Lambda}(4338)$~\cite{Yan:2022wuz,Nakamura:2022jpd}),
$I=0$ $\Xi_c \bar{D}^{(*)}$
and $I=1/2$ $\Sigma_c^{(*)} \bar{D}^{(*)}$ systems, where the new
predictions are presented in Table \ref{tab:predictions-Pcs4338-input}.
The most evident difference with the predictions derived
from the $P_{\psi}^N(4312)$ is the reduced attraction,
which disfavors the formation of bound states
(but favors virtual states).
Curiously, the molecular ratios $R_{\rm mol}'$ with the new input are identical
to the previous ones, where the reason is that the
$\mu_{\rm mol} C^{\rm sat}_{\rm mol}$ of the $P_{\psi}^{N}(4312)$ and
$P_{\psi s}^{\Lambda}(4338)$ differ by only a $0.3\%$, smaller than the accuracy
with which they are listed in Table \ref{tab:predictions-Pcs4338-input}.
From this coincidence it can be inferred that only the molecules for which
$R_{\rm mol} > 1.0$ when the $P_{\psi}^{N}(4312)$ was used as input will survive
as bound states when the $P_{\psi s}^{\Lambda}(4338)$ is the reference state.

It is worth noticing that neither the mass of the $P_{\psi s}^{\Lambda}(4459)$
nor the masses of the $P_{\psi}^N(4312/4440/4457)$ pentaquarks
are reproduced in Table \ref{tab:predictions-Pcs4338-input}.
This either suggests that the $P_{\psi s}^{\Lambda}(4338)$ is not a bound state
or, if it is indeed molecular, that its Breit-Wigner mass does not
correspond with the mass of the pole in $\Xi_c \bar{D}$ scattering.
Nonetheless this is a common occurrence with molecular states
that has been extensively discussed in the literature~\cite{Albaladejo:2015lob,Fernandez-Ramirez:2019koa,Yang:2020nrt}.

If we use instead the $P_{\psi s}^{\Lambda}(4459)$ as the reference state (where
it is described as a $\Xi_c \bar{D}^*$ molecule with $I=0$, yielding
$C^{\rm sat}_{\rm ref} = -0.89^{+0.09}_{-0.06}\,{\rm fm}^2$ at
$\Lambda = 1.0\,{\rm GeV}$), we obtain the molecular
spectrum of Table \ref{tab:predictions-Pcs4459-input}.
In this case the predicted masses of the pentaquarks are usually lighter
than their Breit-Wigner masses, though for the $P_{\psi}^N(4440/4457)$
they are still compatible with the experimental values
within uncertainties.
In contrast, the $P_{\psi s}^{\Lambda}(4338)$ and $P_{\psi}^N(4312)$ overbind:
the incompatibility with the $P_{\psi s}^{\Lambda}(4338)$ is particularly
worrying, as it should have the same potential as the $P_{\psi s}^{\Lambda}(4459)$
(at least, if both are isoscalar molecules).
However here it should be noted that the experimental evidence
for the $P_{\psi s}^{\Lambda}(4459)$ is less conclusive
than for its low mass partner.
Future observations might change its mass, or its single-/double-peak nature,
which in turn will help us to update the predictions in our model.

\section{Conclusions}

We have predicted the molecular hidden-charm pentaquark spectrum within
a compact theoretical description of the baryon-antimeson interaction
in terms of a contact-range S-wave potential.
The couplings of this potential are saturated by the exchange of light-mesons
($\sigma$, $\rho$, $\omega$), where a few renormalization group ideas are
included to properly combine the contributions from mesons
with different ranges.
The couplings thus obtained are unique except for an unknown proportionality
constant, which can be easily determined from a reference or input state.

If we use the $P_{\psi}^N(4312)$ as input (assuming it to be a
$\bar{D} \Sigma_c$ bound state), we are able to approximately reproduce
the masses of the $P_{\psi}^N(4440)$ and $P_{\psi}^N(4457)$
as $J=\tfrac{3}{2}$ and $\tfrac{1}{2}$ $\bar{D}^* \Sigma_c$ bound states,
as well as the theorized narrow $P_{\psi}^N(4380)$~\cite{Liu:2019tjn,Liu:2019zvb,Guo:2019fdo,Du:2019pij}.
The extension of this model to the molecular pentaquarks containing
an antitriplet charmed baryon predicts the existence of
$\bar{D} \Xi_c$ and $\bar{D}^* \Xi_c$  $P_{\psi s}^{\Lambda}$ states
with masses of $4327\,{\rm MeV}$ and $4467\,{\rm MeV}$, respectively,
which might very well correspond with the recently observed
$P_{\psi s}^{\Lambda}(4338)$~\cite{LHCb:2022jad} and
the $P_{\psi s}^{\Lambda}(4459)$~\cite{Aaij:2020gdg}.
That is, the present model is compatible with most of the currently
experimentally observed pentaquarks with the exception of
the $P_{\psi}^N(4337)$~\cite{Aaij:2021august},
the nature of which does not seem to fit within the molecular
picture~\cite{Yan:2021nio,Nakamura:2021dix}.
Yet, caution is advised as it remains unclear whether the $P_{\psi}^N(4457)$ and
$P_{\psi s}^{\Lambda}(4338)$ are actual molecules or not~\cite{Kuang:2020bnk,Burns:2021jlu,Burns:2022uha}.

Besides the previous molecular pentaquarks, we predict a relatively rich
spectrum of new, so far unobserved states.
The eventual experimental observation (or absence) of these states will be able
to confirm (or refute) the present phenomenological model, and thus indirectly
constrain the nature of the currently known pentaquarks.
It is also worth noticing that previous theoretical works
have predicted molecular spectra that are similar to ours~\cite{Liu:2019tjn,Xiao:2019aya,Valderrama:2019chc,Liu:2019zvb,Guo:2019fdo,Xiao:2019gjd,Wang:2019nvm}.
Even though the quantitative details of the spectrum might change by varying
the parameters or by choosing a different input state {(a point that we
  have partially explored for a few selected molecules by using
  the $P_{\psi}^{\Lambda}(4338)$ and $P_{\psi s}^{\Lambda}(4459)$ as
  the reference state)},
there are two important qualitative characteristic of
the predicted spectrum: everything else being equal,
we have that
\begin{itemize}
\item[(i)] the central and spin-spin potentials of molecular hidden-charm
  pentaquarks tend to be more attractive if they are in the smaller-sized
  representations of SU(3)-flavor, i.e. singlets are more attractive
  than octets for $\bar{H}_c T_c$ configurations and, provided
  the evaluation of the spin-spin operator is positive (i.e. meson-baryon
  configurations with the maximum possible spin, $J = J_1 + J_2$),
  octets will also be more attractive than decuplets
  in $\bar{H}_c S_c$ , and
\item[(ii)] for the $\bar{H}_c S_c$ configurations,
  which contain non-trivial dependence on the light-spin of the heavy hadrons,
  higher spin states are  lighter (heavier)
  for the octet (decuplet) configurations.
\end{itemize}
For instance, as a consequence of (i) we expect the most attractive
$\bar{H}_c T_c$ pentaquarks to be the isoscalar $\bar{D}^{(*)} \Xi_c$
configurations, which correspond with the molecular interpretation of
the $P_{\psi s}^{\Lambda}(4338/4459)$.
Meanwhile, from (ii) if the $P_{\psi}^N(4440/4457)$ are $\bar{D}^* \Sigma_c$
bound states, then the $P_{\psi}^N(4440)$ should be $J=\frac{3}{2}$ and
the $P_{\psi}^N(4457)$ $J=\frac{1}{2}$ (because higher spin
implies lighter mass in the present model).
The eventual experimental determination of the spin of the $P_{\psi}^N(4440)$
as well as the discovery of new pentaquarks (particularly decuplets)
would indeed shed light on whether the previous patterns are correct.

\section*{Acknowledgments}

This work is partly supported by the National Natural Science Foundation
of China under Grants No. 11735003, No. 11835015, No. 11975041, No. 12047503, No. 12125507 and No. 12305096,
the Chinese Academy of Sciences under Grant No. XDB34030000,
the Postdoctoral Fellowship Program of the China Postdoctoral Science Foundation under Grant Number GZC20241765, the Fundamental Research Funds for
the Central Universities under Grant No. SWU-KQ25016
and the Thousand Talents Plan for Young Professionals.
M.P.V. would also like to thank the IJCLab of Orsay, where part of
this work has been done, for its long-term hospitality.

%

\end{document}